\documentclass[aps,prc,twocolumn,showpacs,preprintnumbers,nofootinbib,float,superscriptaddress,longbibliography]{revtex4-1}
\usepackage{graphicx}
\usepackage{color}
\usepackage[dvipsnames]{xcolor}
\usepackage{breakurl}
\usepackage[colorlinks=true, pdfstartview=FitV, linkcolor=RedOrange, citecolor=ForestGreen, urlcolor=blue]{hyperref}
\usepackage{amsmath}
\usepackage{amssymb}
\usepackage{bm}
\usepackage{physics}
\usepackage{placeins}
\usepackage{adjustbox}
\usepackage[normalem]{ulem}

\usepackage[maxfloats=256]{morefloats}
\begin{document}
\title{Electromagnetic Radiation from Baryon-Rich Matter in Heavy-Ion Collisions}

\author{Xiang-Yu Wu} \email{xiangyu.wu2@mail.mcgill.ca}
\affiliation{Department of Physics, McGill University, 3600 University Street, Montreal, QC H3A 2T8, Canada}
\author{Charles Gale} \email{charles.gale@mcgill.ca}
\affiliation{Department of Physics, McGill University, 3600 University Street, Montreal, QC H3A 2T8, Canada}
\author{Sangyong Jeon} \email{sangyong.jeon@mcgill.ca}
\affiliation{Department of Physics, McGill University, 3600 University Street, Montreal, QC H3A 2T8, Canada}
\author{Jean-Fran\c{c}ois Paquet} \email{jean-francois.paquet@vanderbilt.edu}
\affiliation{Department of Physics and Astronomy, Vanderbilt University, Nashville, Tennessee 37240, USA}
\author{Bj\"orn Schenke} \email{bschenke@bnl.gov}
\affiliation{Physics Department, Brookhaven National Laboratory, Upton, New York 11973, USA}
\author{Chun Shen} \email{chunshen@wayne.edu}
\affiliation{Department of Physics and Astronomy, Wayne State University, Detroit, Michigan 48201, USA}

\begin{abstract}
We perform a study of electromagnetic radiation in heavy-ion collisions at Relativistic Heavy Ion Collider (RHIC) Beam Energy Scan (BES) and SPS energies using the iEBE-MUSIC framework, which includes 3D dynamical Monte Carlo Glauber initial conditions, MUSIC (3+1)D viscous relativistic hydrodynamics, and the UrQMD hadronic afterburner. The multistage modeling has been calibrated to hadronic data at RHIC-BES energies using a Bayesian analysis. 
Integrating the thermal photon emission rates with the medium evolution, 
we study the direct photon yield and elliptic flow and how they vary with collision energy and emission source. We compare with results obtained by the STAR and PHENIX Collaborations. 
We  employ next-to-leading order thermal QCD dilepton emission rates to compute dilepton invariant mass spectra and extract the effective temperature of the quark-gluon plasma at different collision energies. 
\end{abstract}
\maketitle

\section{Introduction}
\label{sec:intro}
The quark–gluon plasma (QGP) is an exotic state of nuclear matter 
created in relativistic heavy-ion collision facilities such as the Relativistic Heavy Ion Collider (RHIC) and the Large Hadron Collider (LHC). 
The formation of this hot and dense nuclear medium provides a unique opportunity to investigate the many-body properties of quantum chromodynamics (QCD) under extreme conditions. 
Regions of the QCD phase diagram  away from the high $T$ - low $\mu_{\rm B}$ regime were explored experimentally by the RHIC  Beam Energy Scan (BES) program  and previously at the CERN SPS \cite{Satz:2004zd,STAR:2017sal,Luo:2017faz, Bzdak:2019pkr, An:2021wof, Sorensen:2023zkk}. 
Owing to different collision energies and colliding systems, those programs access regions of varying baryon density, and offer insight into the initial structure, evolution, and thermodynamic properties of strongly interacting matter created in heavy-ion collisions~\cite{Braun-Munzinger:2008szb, Fukushima:2013rx, Shen:2020mgh, Monnai:2021kgu, Achenbach:2023pba, Arslandok:2023utm}.

Soft charged hadrons are among the most widely used observables in the study of QGP. Through detailed measurements  of their momentum spectra and azimuthal correlations, it has been shown that the QGP exhibits strong collective flow \cite{Gale:2013da, Shen:2020gef, Heinz:2024jwu,Heinz:2013th,Jeon:2015dfa,Du:2024wjm},
behaving like a viscous fluid with transport coefficients like specific shear and bulk viscosity extracted from final-state hadronic data \cite{Policastro:2001yc, Song:2010mg, Shen:2015msa, Ryu:2015vwa, Schenke:2020mbo}. 
Note however that soft hadrons tend to be  sensitive to the later and colder stages of medium evolution~\cite{Paquet:2017wji}.

Electromagnetic probes such as photons  \cite{Gale:2021emg,Paquet:2015lta,vanHees:2011vb, Shen:2016odt} and dileptons  \cite{Gale:2003iz,Gale:1987ki,vanHees:2007th,Jorge:2025wwp} represent a class of observables that are complementary to strongly interacting ones   \cite{Gale:2025ome,Gale:2003iz,Geurts:2022xmk}. 
They are emitted throughout the entire space-time evolution of the plasma, and because of their minimal interaction with the medium, they escape without significant rescattering. 
This allows photons and dileptons produced in the  medium to preserve undistorted information about the  conditions at their point of emission, such as the local temperature $T(x)$, local flow velocity $u^{\mu}(x) $, and local chemical potential $\mu_B(x).$ 
Although photons and dileptons radiate through the entire evolution of the system, their production rate grows with temperature.
Therefore, photons and dileptons 
are potentially good probes for studying the properties of the early stages. 
Recent studies at LHC energies also suggest that electromagnetic observables can be sensitive to the finite chemical equilibration time \cite{Wu:2024pba, Gale:2021emg, Coquet:2021lca, Wu:2024vyc, Coquet:2023wjk, Shen:2023aeg}.

Compared to that of dileptons, at leading-order, the photon production rates are parametrically enhanced by a factor of $\alpha_s/\alpha_{EM}$ and do not suffer from a large combinatorial background. 
However, the $p_T$-differential photon spectra are affected by blue-shift effects associated with strong collective flow \cite{vanHees:2011vb, Shen:2013vja, Du:2024pbd, Du:2025dot, Massen:2024pnj}, making a temperature extraction more complicated. 
In contrast, although the radial and elliptic flow of the medium Doppler-shifts the transverse-momentum spectra of thermal dileptons, their invariant-mass spectrum is a Lorentz scalar and therefore immune to blue-shift distortions~\cite{Rapp:2014hha, Vujanovic:2022itq}. Therefore, photons and dileptons can provide important complementary information. Dileptons can more directly reflect the medium temperature, while photons are sensitive to both temperature and radial flow and can inform the modeling of the evolving medium. 

By measuring dileptons in different invariant mass windows, one can distinguish contributions from different stages of the plasma.
Specifically, setting aside the dilepton cocktail associated with hadronic and heavy-quark decays,
the low-mass region (LMR, $M < m_\phi $, with $m_\phi$ the mass of the $\phi$-meson) 
is dominated by thermal emission
from the hot hadronic medium,
while the intermediate-mass region (IMR, $m_\phi  < M < m_{J/\psi}$, with $m_{J/\psi}$ the mass of the $J/\psi$-meson) is more sensitive to thermal radiation from the QGP phase \cite{Rapp:1999ej,vanHees:2007th,Rapp:2000pe}.
Therefore, the IMR dilepton spectra provide sensitivity to the high-temperature regions of the QGP medium, and can be used to improve our knowledge of the plasma's highest temperatures~\cite{Churchill:2023zkk, Churchill:2023vpt, Massen:2024pnj}.

By combining signals from soft hadrons, photons, and dileptons -- a multimessenger approach -- one can hope for an eventual robust picture of QGP evolution which includes all the stages preceding hadronic freeze-out. 
In this work, we perform a systematic study of electromagnetic radiation in the baryon-rich region. This includes Au+Au collisions at RHIC BES energies and Pb+Pb/In+In collisions at SPS energies, using the state-of-the-art iEBE-MUSIC framework~\cite{Shen:2017bsr, Shen:2022oyg, Zhao:2022ugy} combined with a recent Bayesian analysis based on RHIC-BES data~\cite{Shen:2023awv, Roch:2024xhh, Jahan:2024wpj, Jahan:2025uye} to select parameters from the Bayesian posterior.

This study introduces several key developments over our previous work~\cite{Gale:2021emg,Paquet:2015lta,Churchill:2023vpt,Churchill:2023zkk}. First, we employ fully dynamical 3D Monte Carlo Glauber initial conditions, as described later in this paper. This approach is particularly critical at RHIC-BES energies, where reduced Lorentz contraction and extended nuclear crossing times make dynamical initialization more physically grounded than the instantaneous approximation commonly adopted in the field. Second, we use thermal photon emission rates~\cite{Shen:2014nfa, Paquet:2015lta,Gervais:2012wd} and NLO thermal partonic dilepton emission rates~\cite{Churchill:2023vpt}  that depend on the baryon chemical potential. We stress that it is important to develop the tools to study the QCD phase diagram at finite $\mu_{\rm B}$,  where lattice QCD only has limited reach~\cite{Nagata:2021ugx}. The tomographic signals provided by EM radiation are therefore very valuable.
Third, this work presents the first calculation of EM radiation at BES energies using (3+1)D hydrodynamic backgrounds systematically constrained by Bayesian calibration against soft hadron data.  This allows for a quantitatively robust modeling of the fireball evolution,  thereby enhancing the predictive power of the EM observables.

This paper is organized as follows: In Sec.\,\ref  {sec:framework}, we briefly introduce the multistage model and present calculations of bulk observables involving soft hadrons. We introduce the photon and dilepton emission sources.
In Section \ref {sec:photon_res} we present results on direct photon production and elliptic flow at finite baryon density. 
Section \ref {sec:dilepton_res} reports on calculations of dilepton invariant mass spectra across the baryon-rich region, and the corresponding effective temperature extracted from these spectra. A discussion of the estimation of Bayesian uncertainty follows, and a summary and an outlook are provided in Sec.\,\ref{sec:conclusion}.

\section{The Theoretical Framework}
\label{sec:framework}
\subsection{Multistage model}
A reliable simulation of electromagnetic ($\gamma$ and $l\Bar{l}$) production at RHIC-BES energies requires a calibrated dynamical model that can describe the evolution of the QGP medium at finite baryon densities. For this purpose, we employ the iEBE-MUSIC framework which comprises an initial state model described in the next paragraph, 
followed by a hydrodynamic evolution provided by the MUSIC viscous relativistic hydrodynamic solver~\cite{Schenke:2010nt, Schenke:2010rr, Paquet:2015lta, Denicol:2018wdp}. Particlization proceeds via the Cooper-Frye~\cite{Cooper:1974mv} prescription with iSS particle sampling~\cite{Huovinen:2012is, Shen:2014vra}. Finally, hadronic rescattering is achieved through the use of UrQMD~\cite{Bass:1998ca, Bleicher:1999xi}. The model parameters are obtained from recently determined posterior distributions from a Bayesian analysis calibrated on RHIC-BES soft hadron observables~\cite{Jahan:2024wpj}. 

The initial stage of heavy-ion collisions is simulated using the 3D Monte Carlo (MC) Glauber model, which combines the standard MC-Glauber approach in the transverse plane with a classical string deceleration model along the longitudinal direction~\cite{Shen:2017bsr, Shen:2017fnn, Shen:2018pty, Shen:2022oyg}. Given the inelastic nucleon-nucleon cross section $\sigma_{\rm NN}$ and the impact parameter $b$, binary collisions between nucleons from the projectile and target nuclei are determined from their transverse coordinates.
Each participating nucleon is further treated as consisting of three hotspots related to the constituent quarks and one hotspot representing the rest of the energy and momentum of the incoming nucleon. 
\begin{figure}[htb]
	\centering
	\includegraphics[width=1\linewidth]{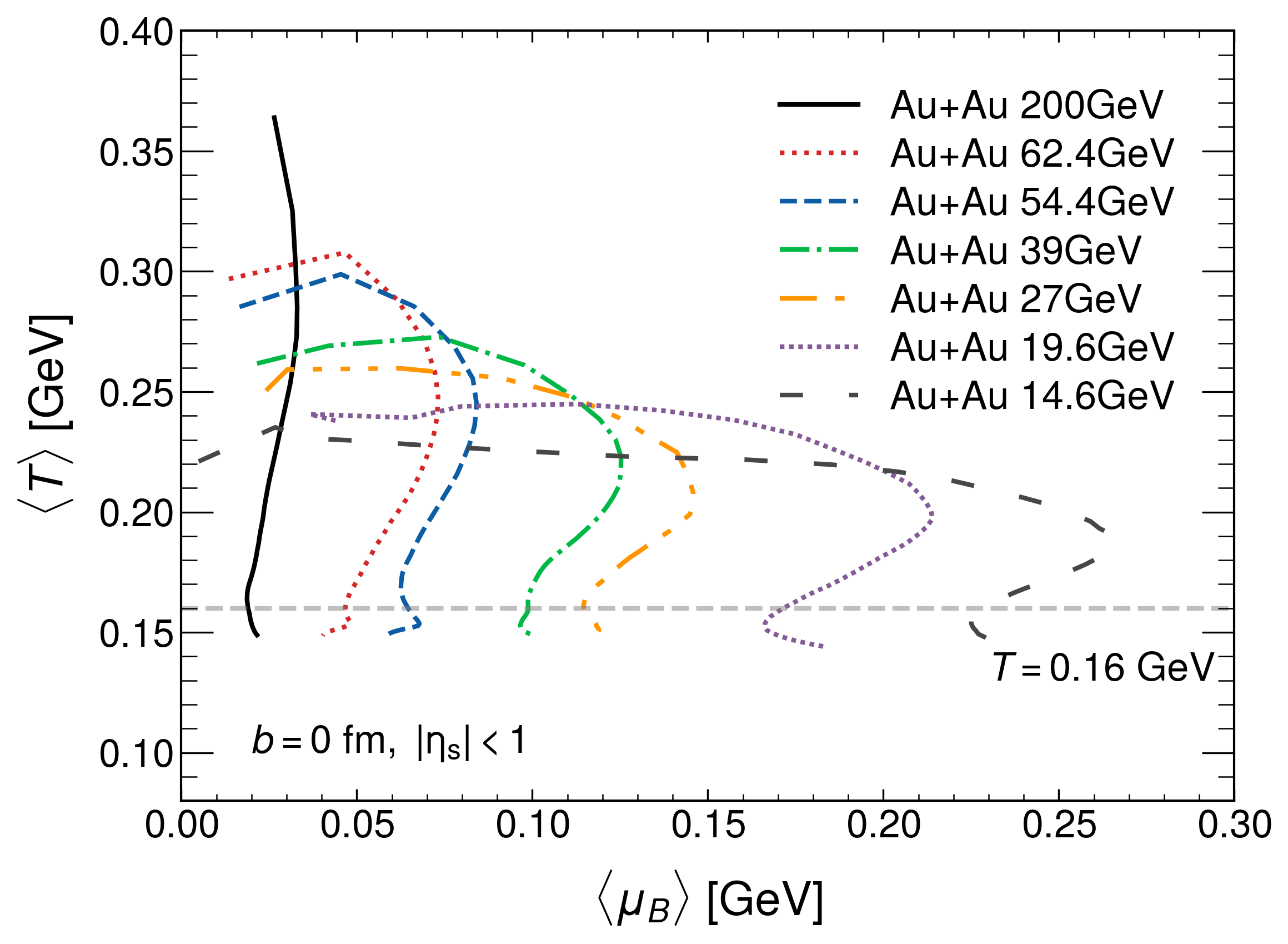}
 \caption{ Hydrodynamic expansion trajectories in a single Au+Au collision with b = 0 and $|\eta_s|< 1$ from 14.6 to 200 GeV. }
	\label{fig:trajectories}
\end{figure}
\begin{figure*}[!htb]
	\centering
	\includegraphics[width=0.8\linewidth]{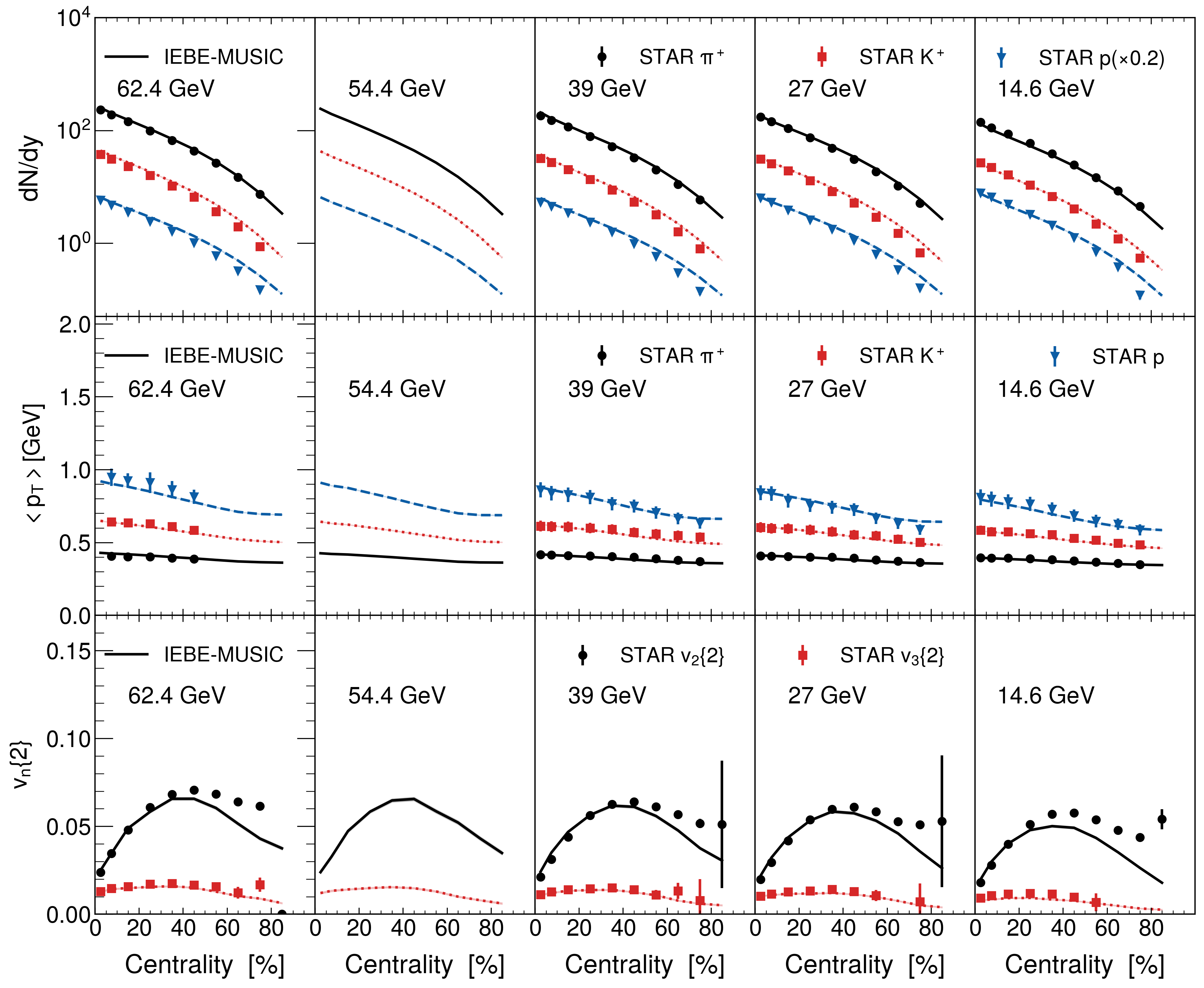}
 \caption{Identified particle yields $dN/dy$, their average transverse momentum $\langle p_T \rangle$, and charged hadron anisotropic flow coefficients $v_n\{2\}$ with $n = 2, 3$ as functions of collision centrality in Au+Au collisions at $\sqrt{s_{\rm NN}}$ = 62.4, 54.4, 39, 27 and 14.6~GeV. Experimental data are taken from the STAR Collaboration \cite{STAR:2019vcp,STAR:2017sal,STAR:2017idk,STAR:2016vqt}. }
	\label{fig:hadrons_14_27_54}
\end{figure*}
After each nucleon–nucleon collision, the wounded partons are connected by strings and experience longitudinal deceleration over {a constant time interval}  $\tau_{\rm{hydro}} = 0.5~\rm{fm}/c$ in the parton collision rest frame~\cite{Shen:2017bsr,Shen:2017fnn}.  The associated energy loss along the beam direction results in a shift in rapidity, which is parametrized using a piecewise function~\cite{Shen:2023awv}. Subsequently, the strings fragment and deposit energy and momentum into the QGP medium.
The incoming baryon charges are either deposited at the string ends or fluctuate toward the string center with a probability $\lambda_B \in [0, 1]$,  motivated by the baryon junction picture~\cite{Kharzeev:1996sq, Pihan:2024lxw}.
To incorporate a more realistic initial longitudinal structure of the system, the 3D-Glauber model introduces a parametrized tilted longitudinal profile for energy deposition, motivated by the observed final-state directed flow~\cite{Bozek:2010bi}. 
Finally, a blast-wave-like initial transverse flow velocity is used to model the transverse expansion effects in the pre-equilibrium stage. The details of the source terms' spatial distributions are specified in Ref.~\cite{Shen:2022oyg}.

\begin{figure}[!htb]
	\centering
	\includegraphics[width=1.0\linewidth]{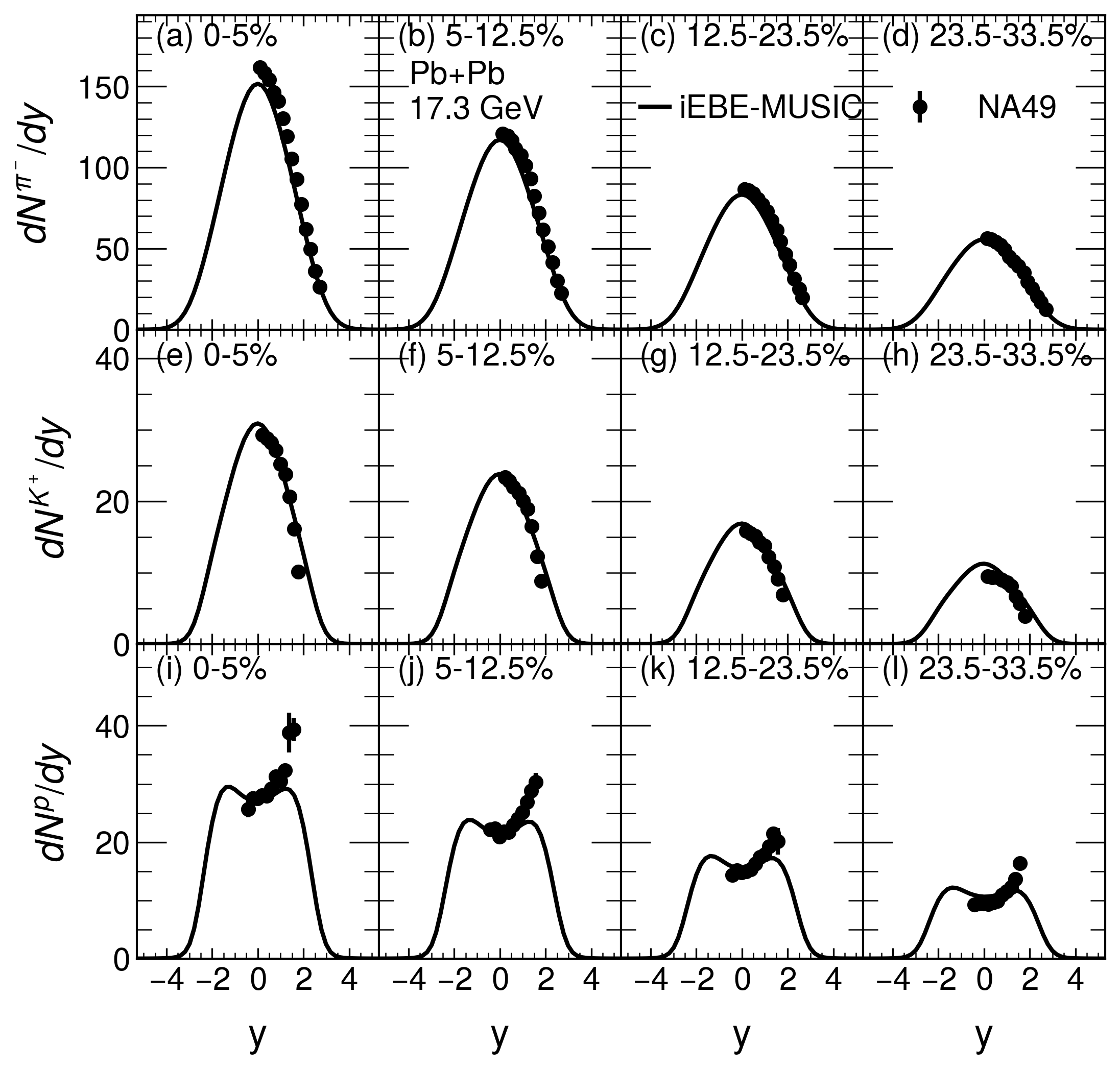}
 \caption{ Rapidity distributions of $\pi^-$ (first row), $K^+$ (second row), and protons (third row) in Pb+Pb collisions at $\sqrt{s_{\rm NN}} = 17.3$ GeV, compared with experimental measurements from the NA49 Collaboration \cite{NA49:2012rsi}. }
	\label{fig:dNdy_sps}
\end{figure}

We emphasize that, in the 3D-Glauber model, the energy, momentum, and net baryon densities are initialized through sources in the hydrodynamic equations~\cite{Shen:2017bsr, Shen:2022oyg}. 
\begin{equation}
\partial_\mu T^{\mu \nu} = J^\nu, \quad \partial_\mu J_B^\mu = \rho_B.
\end{equation}
This is in contrast with initial condition models used at higher collision energy, which initialize hydrodynamics using boundary conditions on a constant-$\tau$ hypersurface. To close the system of equations, we employ the NEOS-BQS crossover equation of state (EOS) \cite{Monnai:2019hkn}, which is based on HotQCD lattice QCD results extended to finite chemical potential by a Taylor expansion \cite{HotQCD:2014kol, HotQCD:2012fhj, Ding:2015fca, Bazavov:2017dus}. The EOS is constructed under the assumption of strangeness neutrality ($n_s = 0$) and a fixed electric charge-to-baryon ratio ($n_Q/n_B = 0.4$), reflecting the typical composition of heavy nuclei such as Au and Pb.

We consider shear and bulk viscous effects in hydrodynamic evolution by solving the second-order Denicol-Niemi-Molnar-Rischke (DNMR) theory~\cite{Denicol:2012cn}. The specific shear viscosity of the QGP is modeled as a function of the net baryon chemical potential but is constant in temperature. The specific bulk viscosity is parametrized as a function of both temperature and chemical potential. For details regarding the parametrizations of shear and bulk viscosities, we refer the reader to \cite{Jahan:2024wpj, Shen:2023awv}.

As the QGP medium expands and the local energy density drops below an energy density $\epsilon_{\rm sw}$, the Cooper-Frye prescription \cite{Cooper:1974mv} is employed to convert fluid elements into hadrons on the local switching hypersurface. Out-of-equilibrium corrections based on Grad's moment method~\cite{Ryu:2015vwa, Schenke:2020mbo, Zhao:2022ugy, Ryu:2023bmx, Roch:2025pcj} are included to map the shear and bulk viscous tensors from fluid cells to the momentum distribution of hadrons.
The thermally produced hadrons further undergo hadronic cascades and resonance decays, which are simulated using the UrQMD afterburner.

Before presenting the final soft-hadron observables, it is also instructive to show the hydrodynamic expansion trajectories, as displayed in Fig.~\ref{fig:trajectories}. The figure presents $\langle T \rangle$ and $\langle \mu_B \rangle$ for a single Au+Au event with $b = 0$ and $|\eta_s| < 1$ from $\sqrt{s_{NN}} = 14.6$ to $200$ GeV under dynamical initialization. Here, we define the average of a quantity $O$ as $\langle O \rangle = \sum O \gamma e / \sum \gamma e$, where the sum runs over all fluid cells with $T > 0.12$ GeV, $e$ is the local energy density, and $\gamma$ is the Lorentz factor. Unlike the case with instantaneous initialization shown in \cite{Churchill:2023vpt,Denicol:2018wdp}, the dynamical initialization model exhibits a much slower decrease of $\langle T \rangle$ at early times, while $\langle \mu_B \rangle$ gradually increases. This behavior reflects the interplay between the expansion of the system and the continuous source deposition. The effect becomes more pronounced at lower collision energies because of the longer overlap time. After the source deposition is completed, both $\langle T \rangle$ and $\langle \mu_B \rangle$ decrease, approximately following a trajectory of constant $s/n_B$.

The model parameters are calibrated using Bayesian inference with the experimentally determined hadron multiplicity $dN/dy$, mean transverse momentum $\langle p_T \rangle$, and charged hadron anisotropic flow $v_n\{2\} (n = 2, 3)$ in Au+Au collisions at 7.7, 19.6, and 200 GeV in the RHIC-BES program, see also \cite{Jahan:2024wpj}.
From the posterior distribution of model parameters, we sample different parameter sets to simulate individual heavy-ion collisions instead of running minimum bias events with one model parameter set. This approach ensures that the obtained ensemble of minimum bias collisions explores as many model parameter sets as possible from the posterior distribution. As a result, the event averages of the final-state observables capture the variations from the posterior distribution, and error bands show those systematic model uncertainty.

Next, we perform model predictions for hadronic observables at several collision energies between 7.7 and 200 GeV based on the approach just described. In Fig.~\ref{fig:hadrons_14_27_54}, we present the model predictions for the centrality dependence of identified particle yields and transverse momentum spectra, as well as the elliptic ($v_2$) and triangular ($v_3$) flow of charged hadrons for $\sqrt{s_{\rm NN}}$ = 62.4, 54.4, 39, 27 and 14.6~GeV Au+Au collisions. Experimental data from the STAR Collaboration are shown for comparison \cite{STAR:2019vcp,STAR:2017sal,STAR:2017idk,STAR:2016vqt}. The results at 54.4~GeV serve as testable predictions for future measurements. We observe that our model predictions describe the STAR measurements at these collision energies, demonstrating that the (3+1)D hybrid model, together with the Bayesian calibration, provides a reliable interpolation for collision energies spanning the RHIC BES program.

Figure~\ref{fig:dNdy_sps} shows the rapidity-dependent hadron production $dN/dy$ for $\pi^-$, $K^+$, and protons in Pb+Pb collisions at $\sqrt{s_{\rm NN}} = 17.3$~GeV measured by the NA49 Collaboration \cite{NA49:2012rsi}. Note that even in this collision system that was not used in the Bayesian analysis, the model provides a very good description of the experimental data for $\pi^-$ and $K^+$ production. For protons, the model describes the data well in the mid-rapidity region but underestimates the data at finite rapidity. This tension may arise from contributions from spectators, not included in the calculations.

Overall, our (3+1)D hybrid framework, combined with the Bayesian posterior distributions, provides a robust and predictive tool for describing the dynamical evolution of the collision systems at RHIC-BES energies. This lays the foundations of our study of thermal photons and dileptons.

\subsection{Electromagnetic radiation}
\label{sec:EM_sources}
Using the dynamical medium calibrated to reproduce measured hadronic observables, the emitted electromagnetic radiation is calculated.
Since EM probes can be generated throughout all stages of heavy-ion collisions, both photon and dilepton signals include contributions from various sources. 
The yields of thermal photons and dileptons can be estimated by integrating the thermal emission rates over the space-time volume $d^4x$, taking into account the local temperature $T(x)$, local baryon chemical potential $\mu_B(x)$, and local flow velocity $u^\mu(x)$ of the strongly interacting  medium. 

\begin{align}
&E \frac{dN_{\gamma}}{d^3p} = \int d^4x \Big( E \frac{d\Gamma_{\gamma}}{d^3p} \big(P^\mu, T(x), \mu_B(x),  \notag \\
&\hspace{11em} u^\mu(x), \pi^{\mu\nu}(x), \Pi(x) \big) \Big) \label{eq:photon_rate}\\
&\frac{dN_{l\bar{l}}}{d^4p} = \int d^4x \left( \frac{d\Gamma_{l\bar{l}}}{d^4p} \left(P^\mu, T(x), \mu_B(x), u^\mu(x) \right) \right) \label{eq:dilepton_rate}
\end{align}
Here, $ E \frac{d\Gamma_{\gamma}}{d^3p} $ and $\frac{d\Gamma_{l\bar{l}}}{d^4p}$ represent the thermal emission rates of photons~\cite{Paquet:2015lta} and dileptons~\cite{Churchill:2023vpt},
 respectively. For the partonic part of the thermal photon rates, leading-order contributions are considered, with shear and bulk viscosity corrections included for the $2\to 2$ contributions~\cite{Shen:2014nfa, Paquet:2015lta,Gervais:2012wd}. The thermal dilepton rates are calculated with finite-temperature field theory \cite{Kapusta:2023eix} up to NLO in $\alpha_s$, without viscosity corrections: a consistent and uniform treatment of those corrections in all EM channels has yet to be completed.  Since the thermal emission rates are defined in the local rest frame (LRF) of the fluid cells, the spectra in Eq.\,(\ref {eq:photon_rate}) and (\ref{eq:dilepton_rate}) implicitly include a Lorentz boost, determined by the local fluid velocity. 
 It is important to emphasize that the emission rates employed in this study explicitly incorporate the dependence on the baryon chemical potential $\mu_B$ through the modifications of quark and antiquark distribution functions, as also of Debye and asymptotic quark masses. Although non-zero $\mu_B$ modifies the thermal dilepton invariant mass spectra~\cite{Churchill:2023vpt}, the net effect remains minimal due to a competition between LO phase-space suppression from antiquark depletion and NLO enhancement driven by $\mu$-dependent thermal screening and the LPM effect. In contrast, finite $\mu_B$ tends to enhance the magnitude of the low $p_T$ portion of the photon spectrum~\cite{Du:2024pbd,Shen2023PhotonBESSlides,Traxler:1994hy}, which also drops faster with increasing momentum, a behavior primarily driven by the increased contribution of quark bremsstrahlung processes in baryon-rich environments~\cite{Gervais:2012wd}.
Importantly, in our treatment of the rates of photon and dilepton radiation, the presence of fermions (quarks and antiquarks) is essential.

In addition to contributions from the QGP medium, the hadronic phase also plays a key role in the production of direct photons and thermal dileptons. 
For photons, several hadronic processes are considered~\cite{Paquet:2015lta,Shen:2015qba}, including meson–meson reactions with viscous corrections~\cite{Dion:2011pp,Shen:2014lye}. The most important channels are $\pi+\rho \to \pi+\gamma$, $\pi+\pi \to \rho+\gamma$, $\pi+K^\ast \to K+\gamma$, $\pi+K \to K^\ast+\gamma$, $\rho+K \to K+\gamma$, and $K^\ast+K \to \pi+\gamma$. meson–meson bremsstrahlung (pion–pion bremsstrahlung) is also included explicitly. Contributions from interactions within the $\pi$–$\rho$–$\omega$ system are taken into account~\cite{Heffernan:2014mla,Holt:2015cda}, including the channels $\pi+\rho \to \omega+\gamma$, $\pi+\omega \to \rho+\gamma$, and $\rho+\omega \to \gamma+\pi$. Baryonic effects are included via the many-body in-medium $\rho$ spectral function, which effectively includes contributions from baryon-baryon and meson-baryon bremsstrahlung processes such as $\pi N \to \pi N\gamma$ and $NN \to NN\gamma$ (where $N$ denoting a nucleon).
However, the hadronic stage contributions for dileptons are not included in this study. Our current focus is on dilepton production in the IMR where the partonic contributions are more important~\cite{NA60:2008ctj,Rapp:1999zw,STAR:2024bpc,PHENIX:2009gyd}.

A transition temperature of $T_{\mathrm{tr}} = 0.16~\mathrm{GeV}$ is employed to separate the contributions from the QGP and the hadronic phase in the calculation of thermal photon and dilepton emission. Between $0.16~\mathrm{GeV}$ and $0.12~\mathrm{GeV}$, the photons from the hadronic phase are included, while hadronic thermal dileptons (which will populate mostly the LMR) are not. The photon emission from the hadronic cascade model is estimated by running the hydrodynamic evolution down to $T_{\mathrm{end}} = 0.12~\mathrm{GeV}$. 
This treatment for hadronic thermal photon has been shown to be a good approximation~\cite{Gotz:2021dco}.
However, it should be noted that both $T_{\rm tr}$ and $T_{\rm end}$ directly affect the relative QGP and hadronic contributions, their choices can be thought of as systematic uncertainty in theoretical  models. 
These parameters should be further constrained by future Bayesian analyses.

Photons and dileptons can also originate from ``cold sources'': sources associated with initial hard parton scattering. Specifically, the prompt photons and the Drell–Yan (DY) dileptons, both calculated using perturbative QCD (pQCD). Previous studies  \cite{Wu:2024pba,Coquet:2021lca,Garcia-Montero:2024lbl} have identified DY dileptons as the dominant source in the high invariant mass region (HMR). Their contribution is subleading compared to thermal dileptons in the IMR, and thus, DY dileptons are not considered in this analysis~\cite{Wu:2024pba,Coquet:2021lca}. 
However, prompt photons constitute an irreducible background when trying to  extract a thermal photon signal from experimental data: their contribution has to be estimated. To do so, we employ the NLO pQCD framework INCNLO \cite{Aurenche:1998gv} corrected for cold nuclear matter effects via the nCTEQ15np \cite{Kovarik:2015cma} nuclear parton distribution functions. The \textsc{BFG-II} fragmentation functions \cite{Bourhis:1997yu} are used to calculate the photon contribution from fragmentation processes. We assume equal factorization, renormalization, and fragmentation scales in this study, i.e., $Q_{\text{fact}} = Q_{\text{ren}} = Q_{\text{frag}} = Q$, where the common scale $Q$ is proportional to the transverse momentum of the photon: $Q = \lambda p_T$. We estimate the theoretical uncertainty in prompt photon production. The method is consistent with that employed in our earlier analysis~\cite{Paquet:2016ulk}.
Finally, to determine prompt photon yields in A+A collisions, we scale the results from INCNLO by the number of binary nucleon-nucleon collisions $N_{\rm binary}$, computed with the MC-Glauber model. 
In this work, we study direct photon production in the low-$p_T$ region in RHIC-BES collisions. The accuracy of NLO pQCD photon calculations is reasonable across broad ranges of photon $p_T$ and $\sqrt{s_{NN}}$~\cite{Aurenche:2006vj}. Nevertheless, calculations are expected to be much less reliable at low $p_T$ and $\sqrt{s_{NN}}$, and in these cases, the pQCD calculations should be seen as a useful baseline rather than an accurate evaluation of prompt photons. We emphasize these limitations at various points in the paper when results are especially dependent on the accuracy of the prompt photon calculation.

\section{Photon production and Photon flow}
\label{sec:photon_res}
\begin{figure}[htb]
	\centering
	\includegraphics[width=1.0\linewidth]{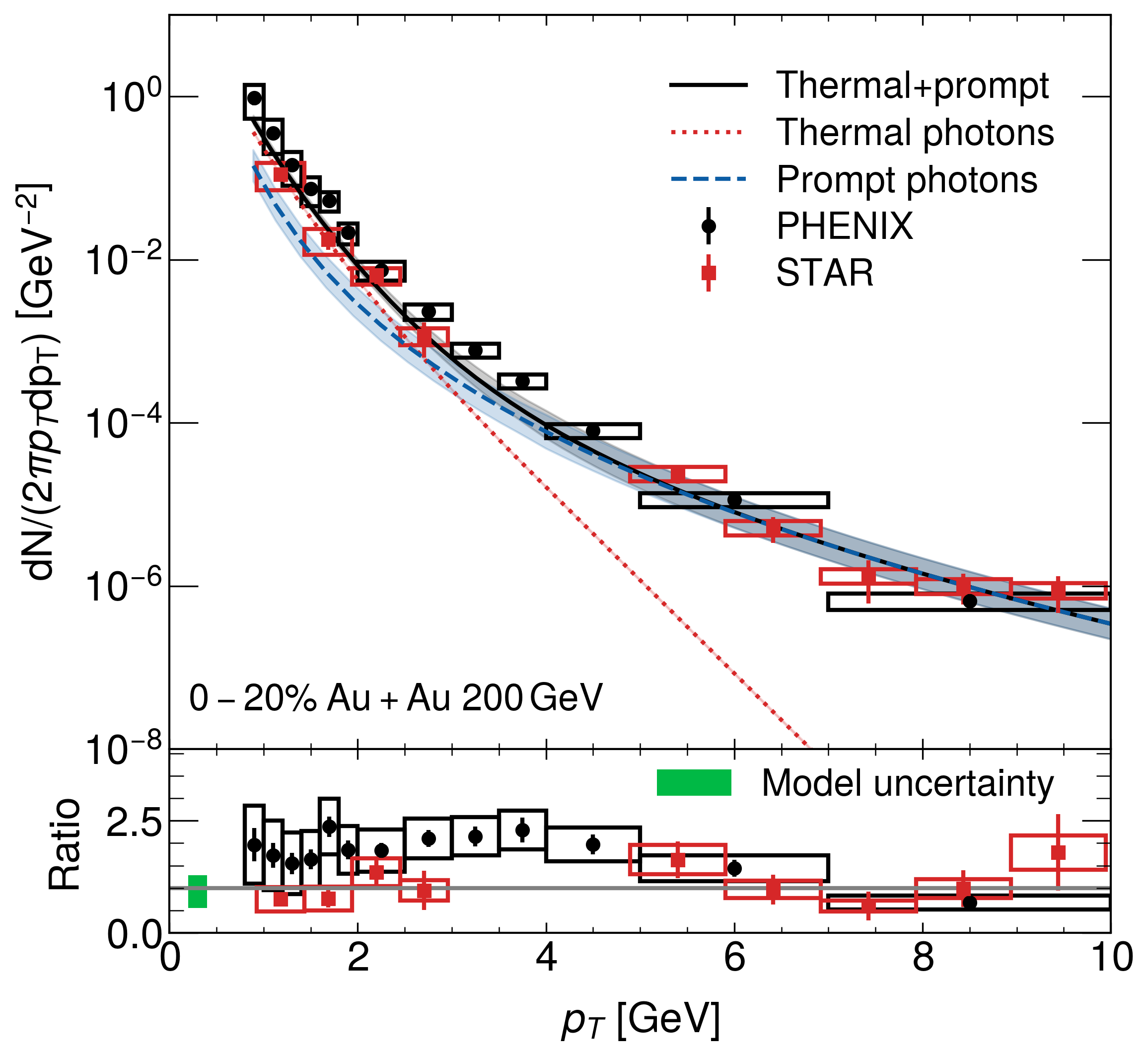}
 \caption{The top panel shows the transverse momentum ($p_T$) distributions of direct photon production and the individual contributions from prompt and thermal photons in 0–20\% central Au+Au collisions at $\sqrt{s_{\rm NN}} = 200$ GeV. Experimental data from the PHENIX \cite{PHENIX:2014nkk} and STAR  \cite{STAR:2016use} Collaborations are included for comparison. The bottom panel shows the ratio of experimental data to the direct photon yield calculated by the model (STAR: red circles with open red boxes; PHENIX: black squares with open black boxes). The green box indicates the averaged relative model uncertainty, including statistical and posterior uncertainties in the direct photon calculations. }
	\label{fig:Directphoton200}
\end{figure}

Figure~\ref{fig:Directphoton200} shows the direct photon production from various sources as a function of transverse momentum $p_T$ at $\sqrt{s_{\rm NN}}$ = 200 GeV in 0-20\% central Au+Au collisions. Experimental data from the STAR and PHENIX collaborations are included for comparison. 
It is worth noting that there is a key difference between our current work and that in Ref.~\cite{Gale:2021emg}: we employ a 3D dynamic MC Glauber model, whereas the previous study used a 2D IP-Glasma initial condition assuming longitudinal boost-invariance with pre-equilibrium evolution. In the dynamic MC Glauber model, the temperature of the system builds up gradually at the early stage, which leads to a reduction of space-time volume for thermal EM radiation at early time compared to the instantaneous proper time that is widely used in conventional hydrodynamic modeling~\cite{Shen:2023aeg}. 
Despite these differences, our results at this energy are qualitatively consistent with those from previous studies~\cite{Gale:2021emg}. In the high-$p_T$ region ($p_T>$  2.5 GeV), prompt photons dominate the direct photon yield,  as they primarily originate from early hard partonic scattering. 
Within scale uncertainties,
the prompt photon yield shows good agreement with STAR and PHENIX data, supporting the validity of applying an $N_{\rm {binary}} $-scaling strategy from pp to AA collisions in the high-$p_T$ regime (particularly for $ p_T > 4 $ GeV). 

The prompt photon contribution becomes subdominant in the low-$p_T$ region ($ p_T < 2.5 $ GeV). Here, thermal photons emitted from the hot and dense fireball are the main source of direct photons, in contrast to the high-$p_T$ region.  
Even though the low-$ p_T $ prompt photon spectrum is calculated based on the NLO pQCD, whose reliability in this regime remains questionable \cite{Paquet:2015lta} and
additional photon production mechanisms, such as photons generated by jet–medium interactions, are not accounted for here \cite{Yazdi:2022cuk}, results from our model still align with the STAR data closely, while underestimating the PHENIX results by a factor of 2–3. This STAR-PHENIX discrepancy is a long-standing and unresolved puzzle~\cite{ David:2019wpt,Gale:2021emg}.

In addition, the lower panel displays the relative uncertainty of the $p_T$-integrated direct-photon yield, which includes both prompt and thermal sources. We find that the prompt contribution dominates the total uncertainty, and the theoretical uncertainty is comparable to the experimental uncertainty at this collision energy. Overall, this result indicates that the prompt component currently sets the precision limit of the direct-photon calculation.

\begin{figure}[htb]
	\centering
	\includegraphics[width=1.0\linewidth]{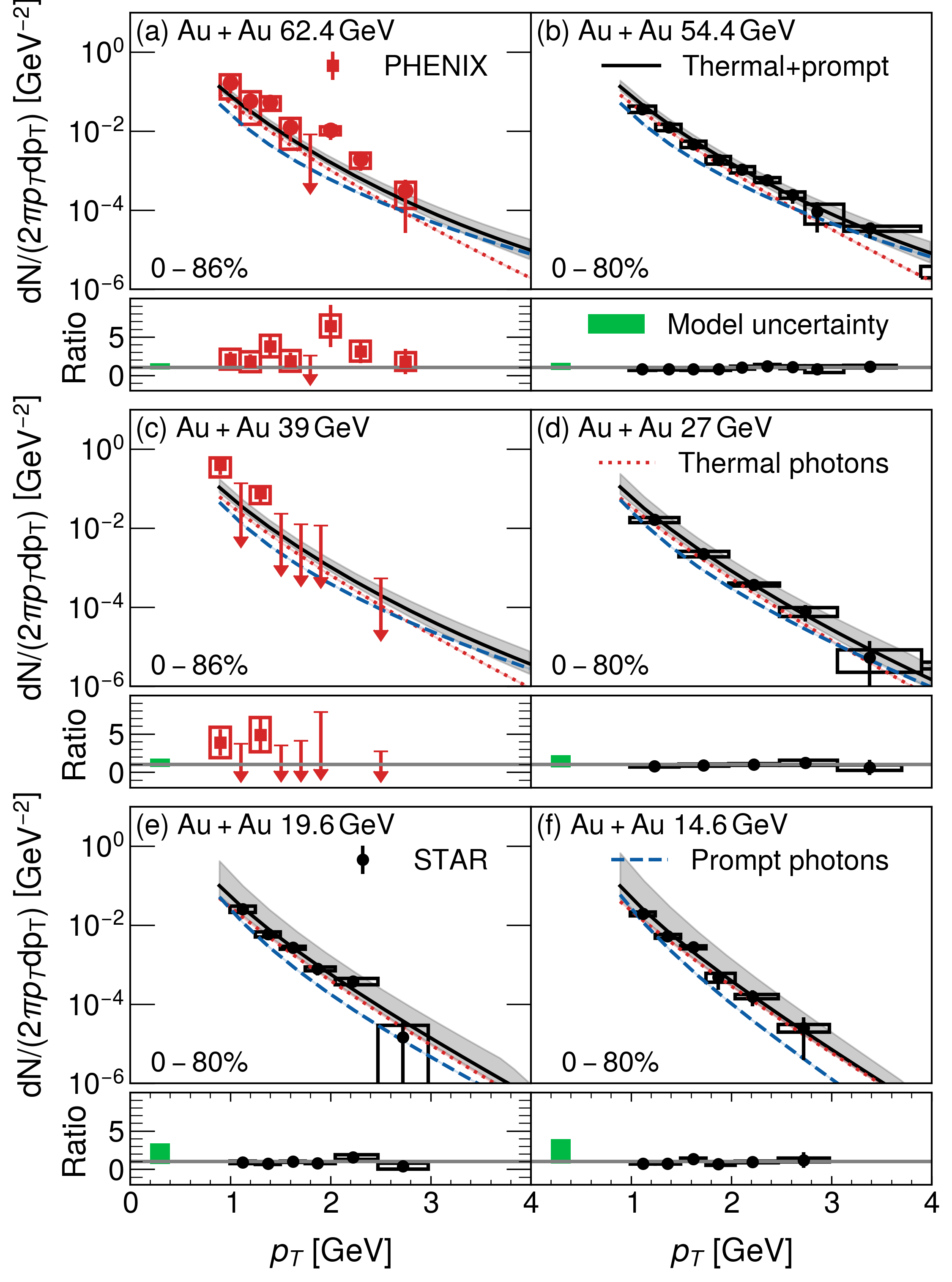}
 \caption{
Transverse momentum ($p_T$) distributions of direct photon production and the individual contributions from prompt and thermal photons in minimum-bias Au+Au collisions at RHIC-BES energies. Experimental data from the STAR and PHENIX Collaborations \cite{xianwenQM,PHENIX:2022qfp} are included for comparison. The direct photon yield (prompt + thermal) is shown as a solid black curve. The prompt and thermal contributions are shown as blue dashed lines and red dotted lines, respectively. Bands indicate the model uncertainties for the direct components. STAR (PHENIX) data points are shown as black circles (red squares). The bottom panel in each subfigure shows the ratio of experimental data to the calculated direct photon yield, with the green box indicating the $p_T$-averaged relative direct photon uncertainty.}
	\label{fig:photons_RHICBES}
\end{figure}

Moving to more recent data, Fig.~\ref{fig:photons_RHICBES} presents the differential transverse momentum spectra of direct photons, highlighting the individual contributions from thermal and prompt photons, at $\sqrt{s_{\rm NN}} =$ 62.4, 54.4, 39, 27, 19.6, and 14.6 GeV in minimum-bias Au+Au collisions. Preliminary STAR data \cite{xianwenQM} and PHENIX measurements~\cite{PHENIX:2022qfp} are included for comparison. 
To improve clarity, Fig.~\ref{fig:photons_RHICBES} shows the prompt photon yield with $Q=\lambda p_T=0.5 p_T$, the thermal photon contribution, and the net direct photon yield with its uncertainty band.
Notably, the direct-photon uncertainty, which is dominated by the scale uncertainty of the prompt component, becomes increasingly significant at lower collision energies.
This highlights once again that the reliability of pQCD-based prompt photon calculations at low $p_T$ and low $\sqrt{s}$ remains limited, and requires careful consideration in future studies. An important approach to address this issue would be to experimentally measure prompt photon production in $p$+$p$ collisions at low energies, which can offer a critical benchmark for validating the theoretical models.

We further observe that our calculation describes the STAR direct photon spectra within experimental uncertainties. 
Similarly to what is shown in Fig.~\ref{fig:Directphoton200}, the model still underestimates the PHENIX data at the lower collision energies.
Moreover, the slope of the direct photon spectrum becomes steeper as the collision energy decreases. This behavior can be explained by a reduction in blue-shift effects and temperature at lower collision energies~\cite{Shen:2012vn}. 
By examining the thermal and prompt photon contributions separately, we find that, on average, the relative contribution of thermal photons to the direct photon yield increases with decreasing collision energy. In particular, below $\sqrt{s_{\rm NN}} = 19.6$ GeV, the thermal photon contribution overtakes that of prompt photons in the $p_T$ range of 1–4 GeV. 
This trend is related to the fact that, even at lower collision energies, where the overlap time between the two nuclei becomes longer (around 2 fm/$c$ at 14.6 GeV), the lifetime of the QGP medium does not decrease significantly. It still lasts for about 10 fm/$c$, providing enough time for thermal photon emission.
In contrast, prompt photon production becomes much more suppressed at low collision energies. This is mainly because lower energies limit the available momentum transfer ($Q^2$) for hard processes, and producing the same $p_T$ prompt photon requires initial partons from larger Bjorken-$x$. Both effects reduce the probability of prompt photon production.

\begin{figure}[tb]
	\centering
	\includegraphics[width=1.0\linewidth]{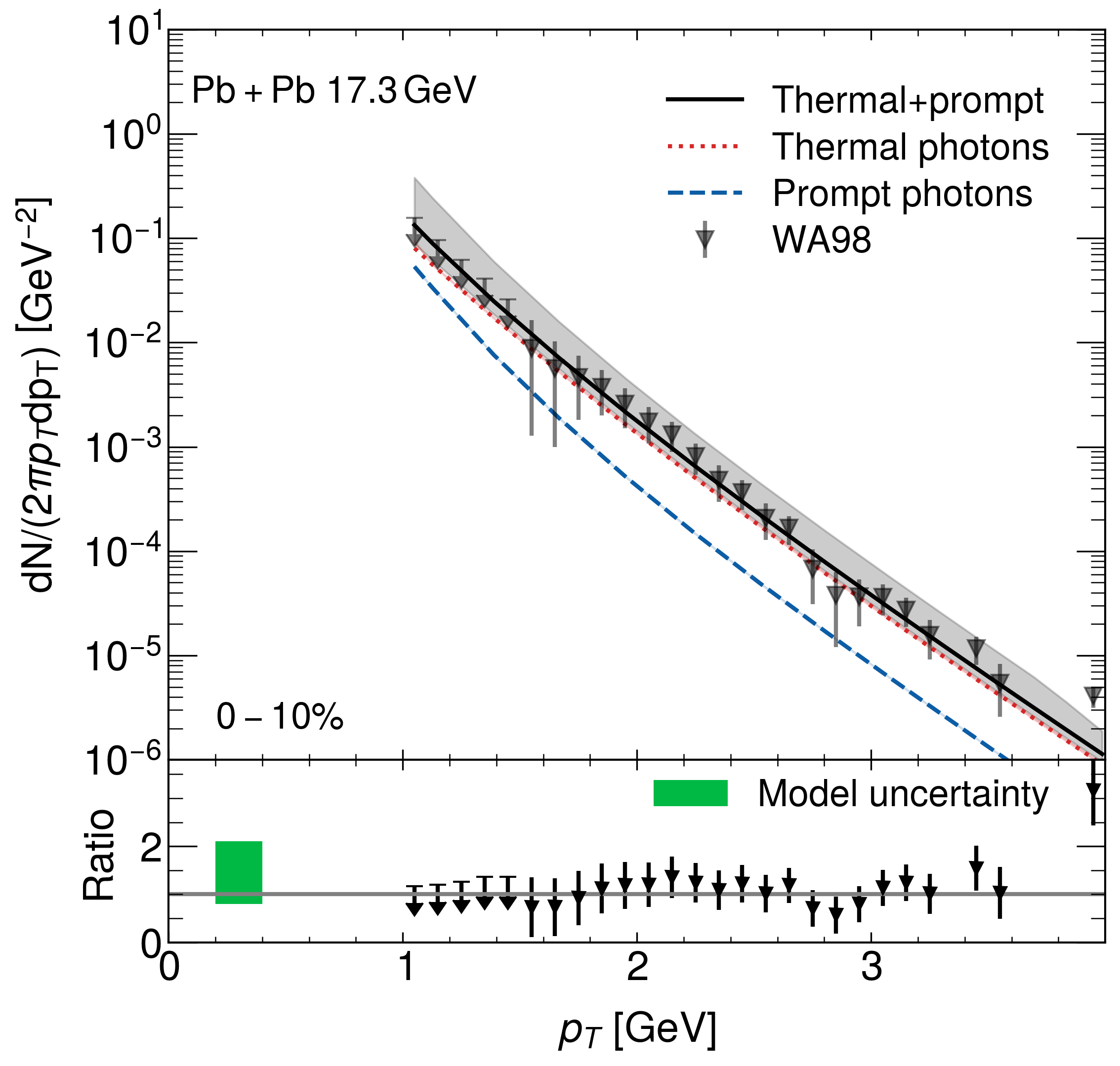}
 \caption{The top panel shows the transverse momentum ($p_T$) distributions of direct photon production and the individual contributions from prompt and thermal photons in 0–10\% central Pb+Pb collisions at $\sqrt{s_{\rm NN}} = 17.3$ GeV. Experimental data from the WA98 Collaboration~\cite{WA98:2000vxl} are included for comparison. The bottom panel shows the ratio of experimental data to the direct photon yield calculated by the model. The green box represents the averaged relative model uncertainty in the direct photon calculations. }
 	\label{fig:photons_WA98}
\end{figure}
We also compute the direct photon production in 0–10\% central Pb+Pb collisions and compare to WA98 measurements in Fig.~\ref{fig:photons_WA98}. This is the first time a (3+1)D viscous hydrodynamic framework combined with Bayesian calibration is employed to compute the direct photon emission from Pb+Pb collisions as measured by the WA98 Collaboration \cite{WA98:2000vxl}. Our model shows good agreement with the data and thermal photons are also found to be the dominant contribution in this Pb+Pb collision system.
By combining the results presented in Figs.~\ref{fig:Directphoton200}, \ref{fig:photons_RHICBES} and \ref{fig:photons_WA98}, it becomes clear that our framework can not only generate results for observables related to charged hadrons, but also for direct photons, and for different collision systems and energies.

\begin{figure}[htb]
	\centering
	\includegraphics[width=1.0\linewidth]{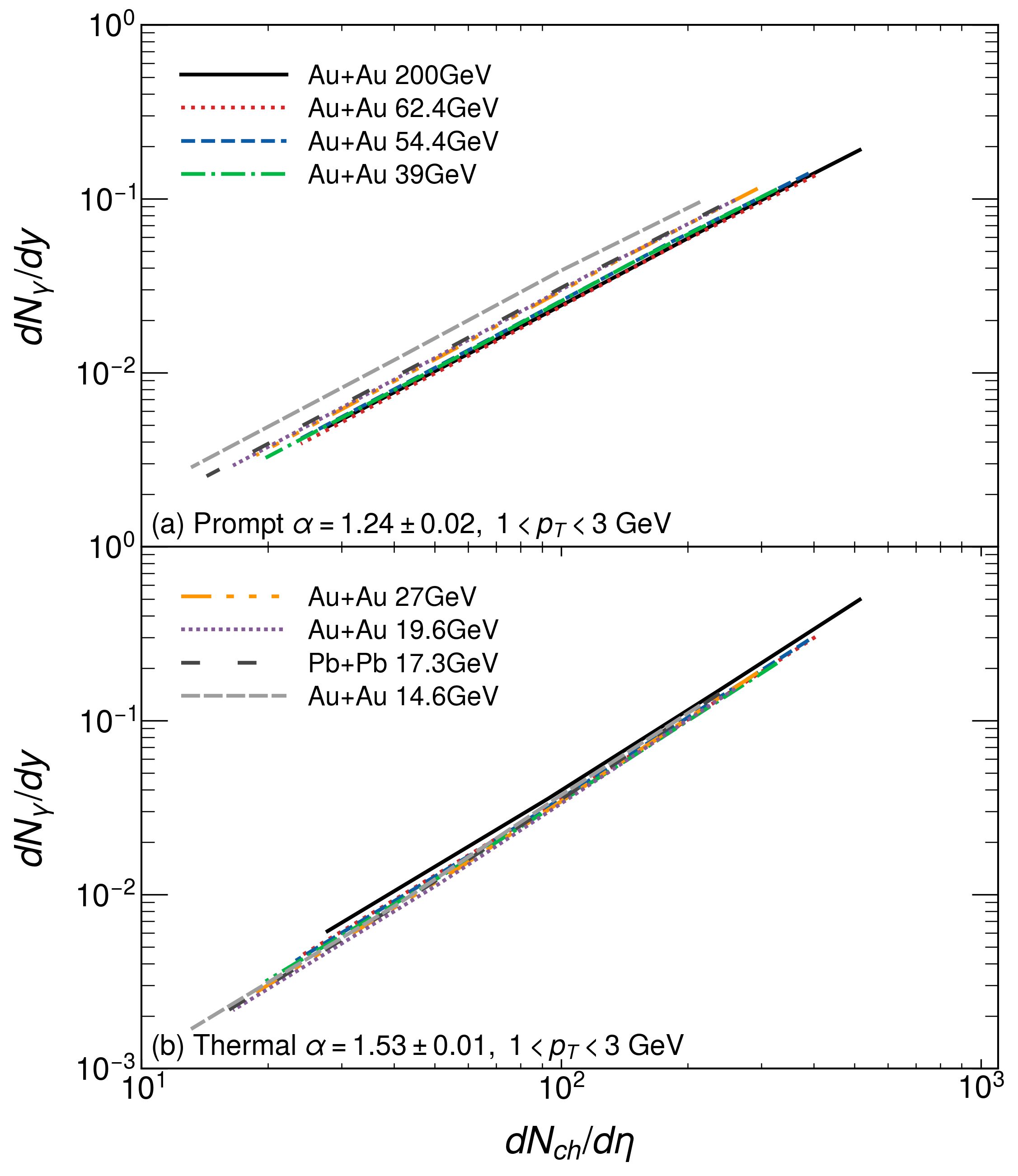}
 \caption{Multiplicity ($dN_{\rm {ch}}/d\eta$) dependence of (a) integrated prompt photon yields and (b) integrated thermal photon yields. Uncertainty bands are not included in this figure.}
	\label{fig:dNgammady_scaling}
\end{figure}

\begin{figure}[htb]
	\centering
	\includegraphics[width=1.0\linewidth]{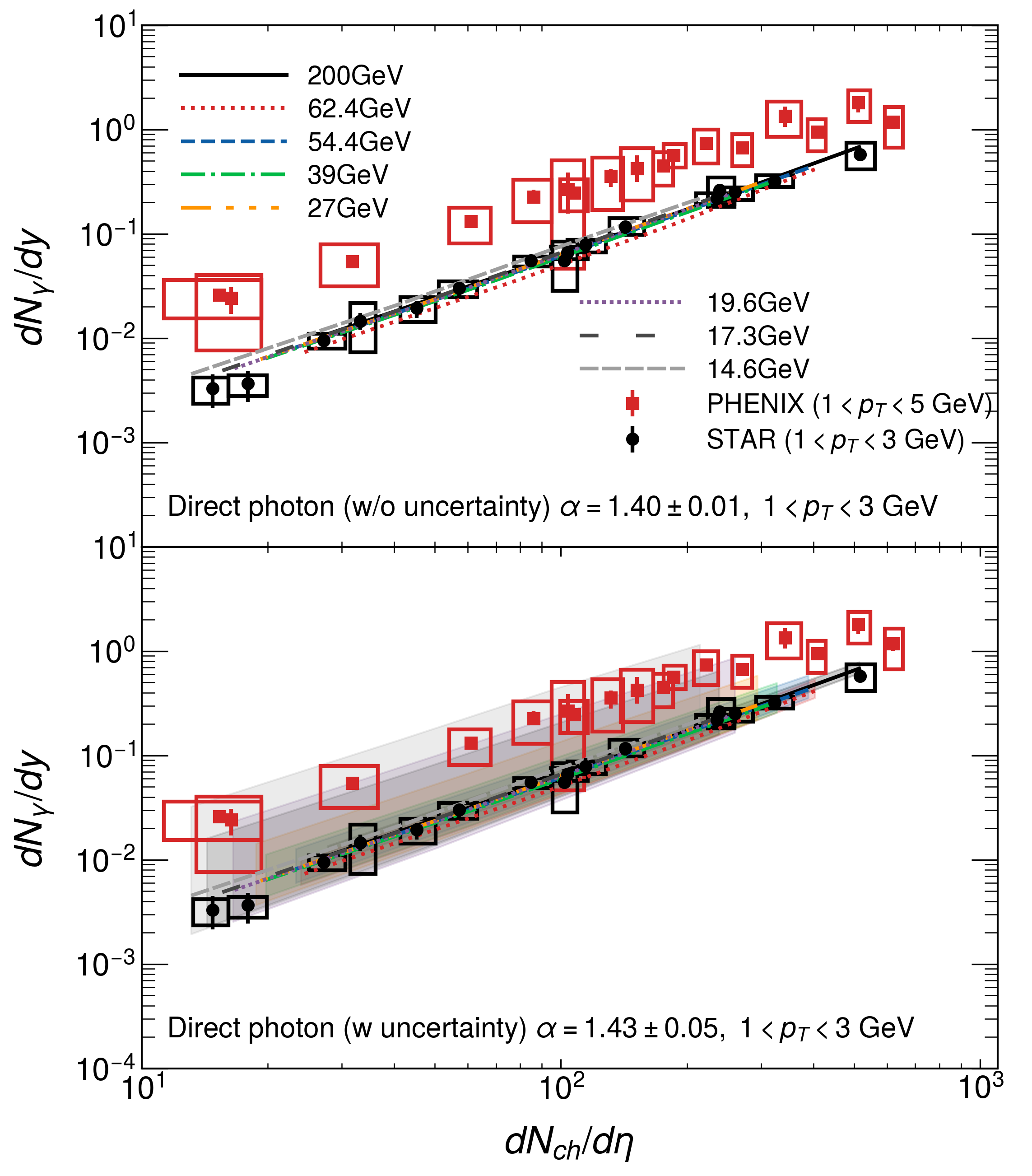}
 \caption{Integrated direct photon yields as a function of multiplicity ($dN_{\rm{ch}}/d\eta$). The upper panel shows results without uncertainties, and the lower panel includes them. Data from the STAR~\cite{xianwenQM} and PHENIX~\cite{PHENIX:2022rsx, PHENIX:2018for} Collaborations are shown for comparison.}
    \label{fig:dNgammady_scaling_total}
\end{figure}

Additionally, we investigate the scaling behavior of the $p_T$-integrated direct photon yield with respect to the charged particle multiplicity ($dN_{\rm {ch}}/d\eta$) as shown in Figs.~\ref{fig:dNgammady_scaling} and \ref{fig:dNgammady_scaling_total}. This analysis includes Au+Au collisions from $\sqrt{s_{\rm NN}} =$ 14.6 to 200 GeV, as well as Pb+Pb collisions at 17.3 GeV.

First, we examine the multiplicity scaling of individual photon sources in Fig.~\ref{fig:dNgammady_scaling}. Uncertainty bands are not included in this figure for clarity.
As a function of centrality, prompt photons are assumed to scale strictly with the number of binary nucleon collisions, which can be related approximately by a power law to the hadron multiplicity~\cite{PHENIX:2018for}. Hence, the scaling of the prompt photons is well understood~\cite{Gale:2019abf}: for each center-of-mass energy, calculations individually scale with the number of binary collisions.
In addition, thermal photons exhibit a similar $(dN_{\rm{ch}}/d\eta)^\alpha$ scaling behavior, but with a larger exponent. Unlike prompt photons, their yields at different collision energies align along a single curve. 
This can be broadly understood from the scaling of the thermal photon yields with the space-time volumes of the collision systems, which are strongly correlated with the system total entropy and charged hadron multiplicity.

After applying a power-law fit of the form $dN^{\gamma}/dy \propto (dN_{\text{ch}}/d\eta)^{\alpha}$, we find that the scaling exponent for integrated prompt photons is $\alpha = 1.24\pm0.02$, which is smaller than that for thermal photons, $\alpha = 1.53\pm0.01$. 
The scaling of the integrated prompt photons, which originates from the relationship between $N_{\text{binary}}$ and $dN_{\text{ch}}/d\eta$, is consistent with those from the PHENIX study     \cite{PHENIX:2018for}, which found that $N_{\text{coll}} \propto (dN_{\text{ch}}/d\eta)^{\alpha}$ with $\alpha = 1.25 \pm 0.02$.

Figure \ref{fig:dNgammady_scaling_total} shows the direct photon yields, including cases with and without uncertainty bands, obtained as the sum of the prompt and thermal components.
Here, we compare direct photon results to experimental data from the PHENIX and STAR Collaborations. 
The observed tension between the PHENIX~\cite{PHENIX:2022rsx} and STAR~\cite{xianwenQM} measurements is consistent with that seen in the $p_T$-differential spectra shown in Figs.~\ref{fig:Directphoton200} and \ref{fig:photons_RHICBES}. Without including uncertainty bands, our model underestimates the direct photon yield compared to the PHENIX data, but shows better agreement with the STAR results. 
However, once the theoretical uncertainty bands are included, the model calculations become compatible with both experimental datasets. 
The scaling exponent $\alpha$ extracted from our model, for the sum of prompt and thermal photons, is $1.40 \pm 0.01$ without including uncertainty bands, and $1.43 \pm 0.05$ when uncertainty is considered. These values are reasonably consistent with the STAR preliminary result of $\alpha = 1.43 \pm 0.04~(\rm {stat}) \pm 0.04~(\rm {sys})$, and higher than the PHENIX value of $\alpha = 1.11 \pm 0.02~(\rm {stat})^{+0.09}_{-0.08}~(\rm {sys})$~\cite{PHENIX:2022rsx}.
Although the scaling exponent $\alpha$ remains similar for direct photon results with and without the uncertainty bands, it should be noted that the magnitude of the integrated direct photon yield still has large uncertainties, which primarily originate from the uncertainties of the prompt photon component. This further highlights the importance of prompt photon measurements in $p+p$ collisions at the RHIC-BES energy regime.
\begin{figure}[tb]
	\centering
	\includegraphics[width=1.0\linewidth]{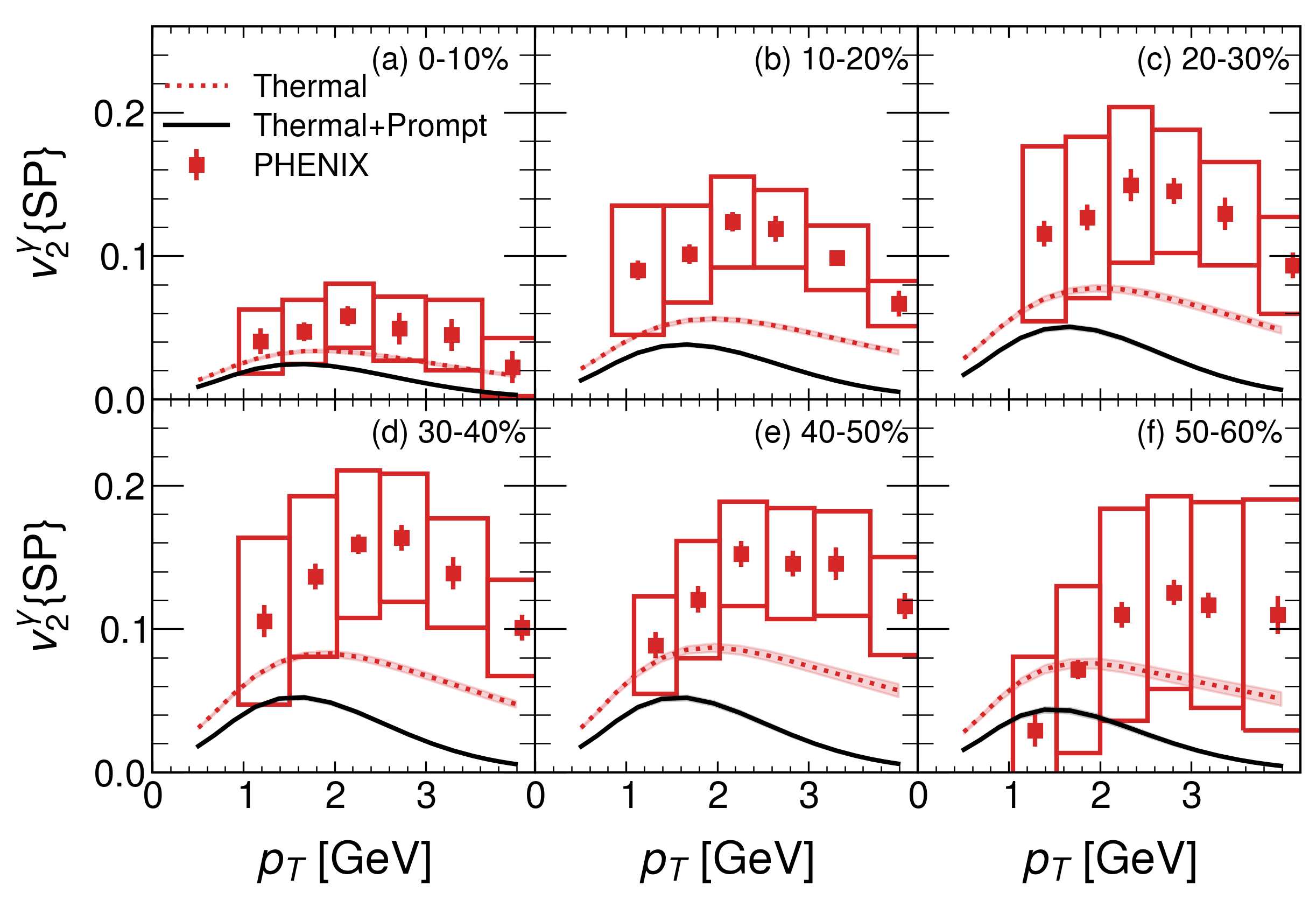}
     \caption{ Transverse momentum ($p_T$) dependence of the direct photon flow and thermal photon flow ($v_2$) in Au+Au collisions at $\sqrt{s_{\rm NN}}$ = 200~GeV for centrality classes from 0–10\% to 50–60\%. The experimental data is taken from PHENIX\cite{PHENIX:2025ejr}.
}
	\label{fig:photon_v2_thermal_propt_cent}
\end{figure}

In addition to the direct photon spectrum, the anisotropic flow of direct photons is another key observable for probing the early-time dynamics of the plasma~\cite{Gale:2020xlg, Gale:2020dum}. It can be calculated via the scalar product method as described in \cite{Gale:2021emg,Paquet:2015lta,PHENIX:2025ejr}, while the prompt-photon contribution is computed with a fixed scale factor $Q=\lambda p_T=0.5 p_T$.
In Fig.~\ref{fig:photon_v2_thermal_propt_cent}, we present the $p_T$-differential elliptic flow ($v^\gamma_2\{\mathrm{SP}\}$) of both direct and thermal photons from central to peripheral Au+Au collisions at 200 GeV. The data is taken from the latest PHENIX analysis \cite{PHENIX:2025ejr}.
Thermal photons exhibit a sizable elliptic flow.  However, when the prompt photon contribution is included,  the overall direct photon elliptic flow $v_2$ is suppressed, particularly in the region $p_T > 2$ GeV. This is because the prompt photons are a large contribution at high $p_T$ and carry negligible anisotropy because of their early production.

As mentioned previously, we use a dynamical initialization that spreads out the energy deposition in time during the early stages of the collisions. This setup reduces the space-time volume for thermal EM emission at early time, which leads to an enhancement in the thermal photon $v_2$~\cite{Shen:2023aeg}.
Nevertheless, even with such a dynamical setup, a noticeable discrepancy remains between the model calculations and the PHENIX data. Fig.~\ref{fig:photon_v2_thermal_propt_cent} shows that, at present, including a dynamic initial condition alone is insufficient to resolve the ``direct photon $v_2$ puzzle'' \cite{David:2019wpt}. 
\begin{figure}[htb]
	\centering
	\includegraphics[width=1.0\linewidth]
    {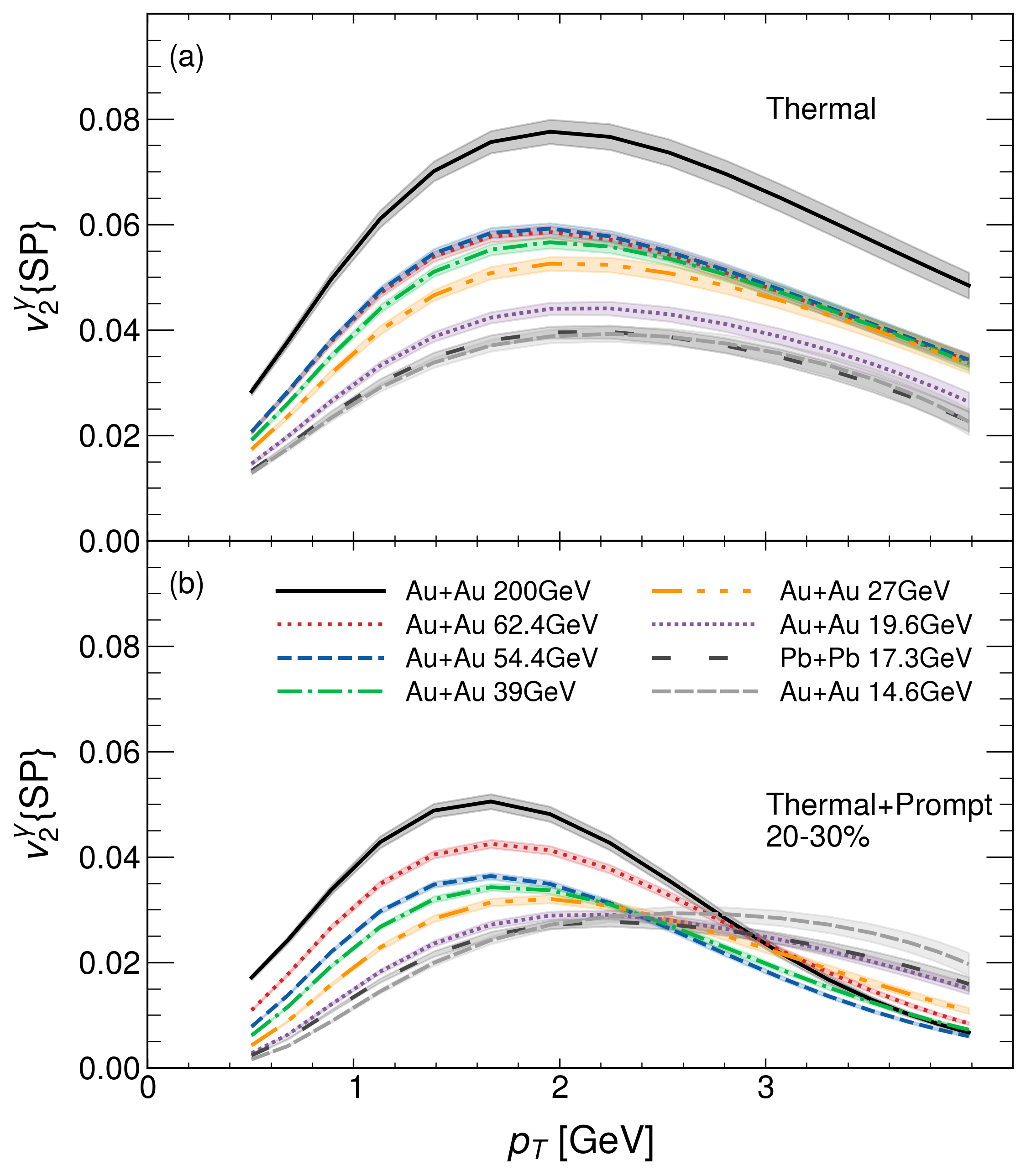}
 \caption{Thermal photon (panel (a)) and direct photon (panel (b)) elliptic flow ($v_2\{\mathrm{SP}\}$) in the 20–30\% centrality class for Au+Au collisions from $\sqrt{s_{\rm NN}}$ = 14.6 to 200 GeV, and for Pb+Pb collisions at $\sqrt{s_{\rm NN}} = 17.3$ GeV.}
	\label{fig:photon_snn}
\end{figure}

\begin{figure*}[htb]
	\centering
	\includegraphics[width=0.8\linewidth]{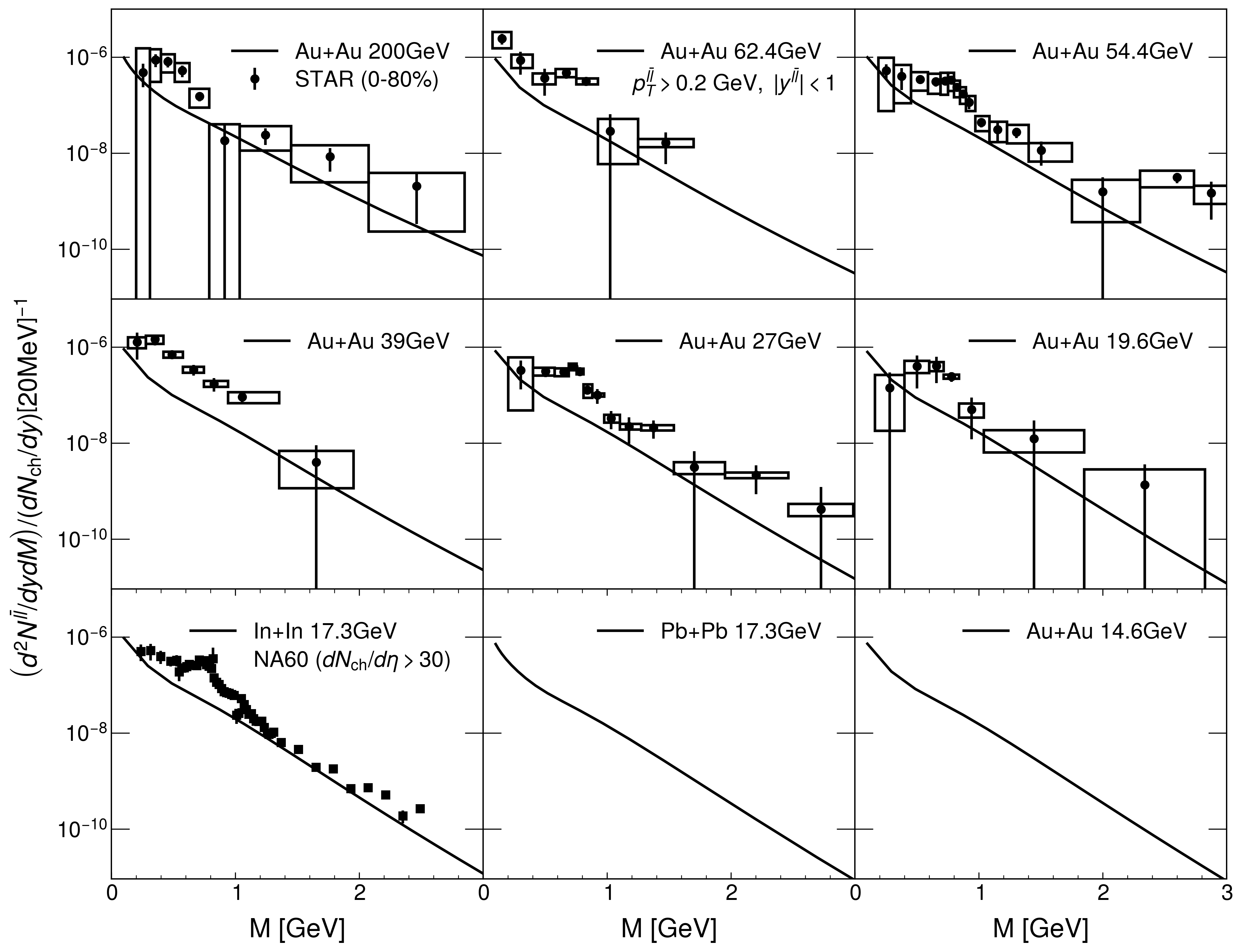 }
 \caption{Dilepton invariant mass excess spectra at mid-rapidity ($|y| < 1$) in Au+Au collisions (0–80\% centrality class) at $\sqrt{s_{\rm NN}}$ = 14.6–200 GeV at RHIC, as well as in In+In collisions ($dN_{\rm {ch}}/d\eta > 30$) and  Pb+Pb collisions (0–80\% centrality class)  at $\sqrt{s_{\rm NN}} = 17.3$ GeV at the SPS. Experimental data are taken from the STAR \cite{STAR:2013pwb,STAR:2015zal,STAR:2023wta,STAR:2024bpc} and NA60 \cite{NA60:2008dcb} Collaborations. }
	\label{fig:Thermaldilepton_norm_00}
\end{figure*}

The collision energy dependence of thermal photon and direct photon elliptic flow ($v_2\{\mathrm{SP}\}$) in the 20–30\% centrality bin is shown in Fig.~\ref{fig:photon_snn}. 
It shows a clear collision-energy hierarchy across the entire
$p_T$ range in the thermal photon $v_2$ (panel (a)) with the values increasing as the collision energy increases.
This behavior is primarily due to the longer lifetime of the QGP medium at higher energies, which allows for more momentum anisotropy to develop in the system and a longer emission duration of thermal photons.
After including the contribution of prompt photons (panel (b)), direct photon $v_2$ with $p_T < 2.5$ GeV still displays this energy hierarchy but has a smaller magnitude due to the suppression from prompt photons. Nevertheless, even at the lowest studied collision energies, the direct photon anisotropic flow is sizable.  In contrast to thermal photons, our simulations suggest that the direct photon $v_2$ with $p_T > 2.5$ GeV increases as the collision energy decreases. This opposite dependence on $\sqrt{s_{\rm NN}}$ is because the dilution from prompt photons reduces at low collision energies~\cite{Gale:2019abf}, according to our current calculation of prompt photons, which as discussed previously do have significant uncertainties. The thermal photons become the dominant source of direct photons with $p_T \in [2.5, 4]$ GeV at low collision energies. Therefore, measuring direct photon flow at low collision energies offers two main benefits. It helps constrain phenomenological models, particularly those related to initial conditions and pre-equilibrium dynamics in the baryon-rich region, and provides a method to probe QGP transport properties at high temperature and large $\mu_B$.

\section{Dilepton production}
\label{sec:dilepton_res}
In this section, we  explore another type of electromagnetic probe: dileptons. 
Figure~\ref{fig:Thermaldilepton_norm_00} shows the dilepton invariant-mass spectra at mid-rapidity in minimum-bias Au+Au collisions at $\sqrt{s_{\rm NN}} = 14.6$–200 GeV at RHIC, as well as in In+In and Pb+Pb collisions at 17.3 GeV at the SPS. For In+In, dileptons are measured as dimuons, while in all other systems they are measured as dielectrons.
These ``excess spectra'' (the cocktail contributions have been subtracted from the inclusive dilepton measurement)  are normalized by the charged particle multiplicity ($dN_{\text{ch}}/dy$).\footnote{Here, we consider charged particle multiplicity $dN_{\text{ch}}/dy$ as the $dN_{\text{ch}}/dy$ sum of $\pi^{\pm}$,$K^{\pm}$, $p$ and $\bar{p}$ for STAR measurements.} 
Here the cocktail denotes the contribution from hadronic sources, including Dalitz decays, the decays of light vector mesons, semi-leptonic decays of open heavy flavor hadrons,  the decays of quarkonium as well as Drell-Yan production.
The detailed cocktail table is referred to in \cite{STAR:2013pwb,STAR:2015zal,STAR:2023wta,STAR:2024bpc}.
We find that our model calculations agree qualitatively with the data from STAR~\cite{STAR:2013pwb,STAR:2015zal,STAR:2023wta,STAR:2024bpc} and NA60~\cite{NA60:2008dcb} in the IMR ($1 < M < 3$\,GeV) within experimental uncertainties, although an underestimation is observed for some collision energies (see Fig.~\ref{fig:Thermaldilepton_norm_00}). Generally, thermal dileptons in this mass range are understood to originate mainly from the QGP phase, and one possible reason for the underestimation is the lack of contributions from the pre-equilibrium stage, as discussed later in this section. It is important to note that thermal radiation from in-medium $\rho$, while essential in the LMR, is known to drop off rapidly above $M \gtrsim 1$ GeV and to become subleading in the IMR, as confirmed in several previous studies~\cite{vanHees:2007th,Ruppert:2007cr,Santini:2011zw,Jorge:2025wwp,Vujanovic:2019yih}. In this study we do not include the sources known to be important in the LMR ($M \lesssim 1$ GeV)~\cite{Rapp:1999ej}, and our results underestimate the data there. Work is ongoing to include all known sources consistently in order to obtain a true excitation function for dileptons, over a wide range of invariant masses.

\begin{figure}[htb]
	\centering
	\includegraphics[width=1.0\linewidth]{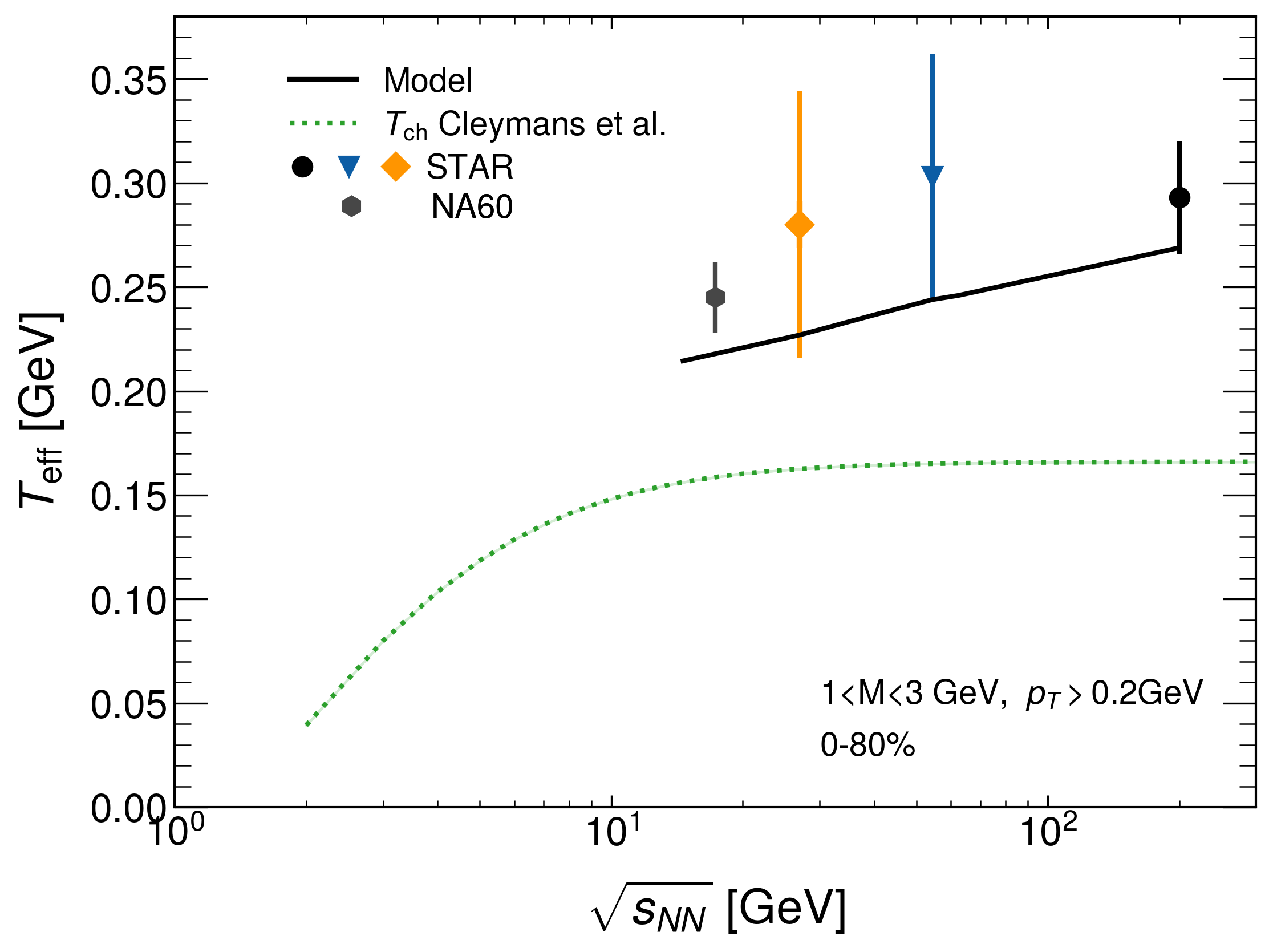 }
 \caption{Effective temperature ($T_{\rm{eff}}$) extracted from dilepton invariant mass excess spectra, compared with measurements from the STAR and NA60 Collaborations \cite{STAR:2024bpc,NA60:2008dcb}. The chemical freeze-out temperatures are taken from the Cleymans \cite{Cleymans:2005xv} for reference.
 }
	\label{fig:Teff_00-80_snn}
\end{figure}
The invariant mass dependence of dileptons in the IMR can be used to define an  effective temperature,
which is extracted by fitting $T$ in Eq.~\ref{eq:T_def} to the dilepton spectrum. In this region, the dilepton thermal rate can be approximated as \cite{Rapp:2014hha}
\begin{equation}
\frac{dR_{l\bar{l}}}{dM} \propto (MT)^{3/2} \;\; {\rm exp}(-M/T).
\label{eq:T_def}
\end{equation}
Previous studies \cite{Churchill:2023vpt,Massen:2024pnj} have shown that the temperature extracted using this functional form is related to the average temperature of fluid cells in hydrodynamic simulations.
Based on this relation, we extract the effective temperature ($T_{\rm eff}$) from the excess invariant mass spectra shown in Fig.~\ref{fig:Thermaldilepton_norm_00}, focusing on the IMR. The extracted values are presented in Fig.~\ref{fig:Teff_00-80_snn}, along with experimental data from the STAR and NA60 Collaborations. 
For comparison, we also include the chemical freeze-out temperatures determined from the Cleymans freeze-out line \cite{Cleymans:2005xv}.

Our model predicts a mild increase of $T_{\rm eff}$ with collision energy, reflecting the hotter and longer-lived QGP medium at higher $\sqrt{s_{\rm NN}}$. Notably, the extracted $T_{\rm eff}$ values are significantly higher than the chemical freeze-out temperature and this is consistent with the understanding that $T_{\rm eff}$ extracted from the IMR represents an average temperature of the QGP phase. 

Although the model results are in agreement with the data within uncertainties, they tend to lie near the lower edge of the experimental error bars. This slight underestimation may originate from the following sources. First, our current calculation does not include the contribution from thermal dilepton emission during the pre-equilibrium stage, particularly before the strings are thermalized. Second, the thermal dilepton emission rate used in the model is based on an ideal equilibrium rate, without including non-equilibrium corrections, which may also lead to an underestimation of the effective temperature~\cite{Ryblewski:2015hea, Vujanovic:2017psb, Vujanovic:2019yih}. Third, in the mass region exceeding the $\phi$ meson, continuum multi-meson states (e.g., 4$\pi$ states) may be generated and yield a small and non-negligible contribution to the total thermal dilepton spectrum~\cite{vanHees:2007th}.

To further explore the impact of the pre-equilibrium stage on dilepton production, we compare three initialization scenarios at the top SPS energy. The case of dynamical deposition and two instantaneous deposition cases with initial proper times $\tau_0$ = 0.5 fm/$c$ and $\tau_0=1.5$ fm/$c$ are considered. 
In the instantaneous deposition case, we neglect the finite longitudinal thickness of the incoming nuclei and assume all binary collisions and string production are taken to occur on the $z = 0$~fm plane at $\tau = 0$~fm/$c$. Then all the strings decelerate by a constant longitudinal proper time ($\Delta\tau = 0.5$~fm/$c$ or $1.5$~fm/$c$ in this case) before they are deposited into hydrodynamic fields as source terms~\cite{Shen:2023aeg}. Subsequently, all the energy and baryon number are simultaneously injected into the QGP medium at a fixed longitudinal proper time $\tau_0$. At $\sqrt{s_\mathrm{NN}} = 17.3$\,GeV, the choice of the two instantaneous deposition cases at $\tau_0 = 0.5$~fm/$c$ and $1.5$~fm/$c$ sandwich the dynamical deposition case in which the QGP medium gradually appears between these two time stamps. A previous study \cite{Shen:2023aeg} showed that hadronic observables were insensitive to these three cases.

\begin{figure}[tb]
	\centering
	\includegraphics[width=1.0\linewidth]{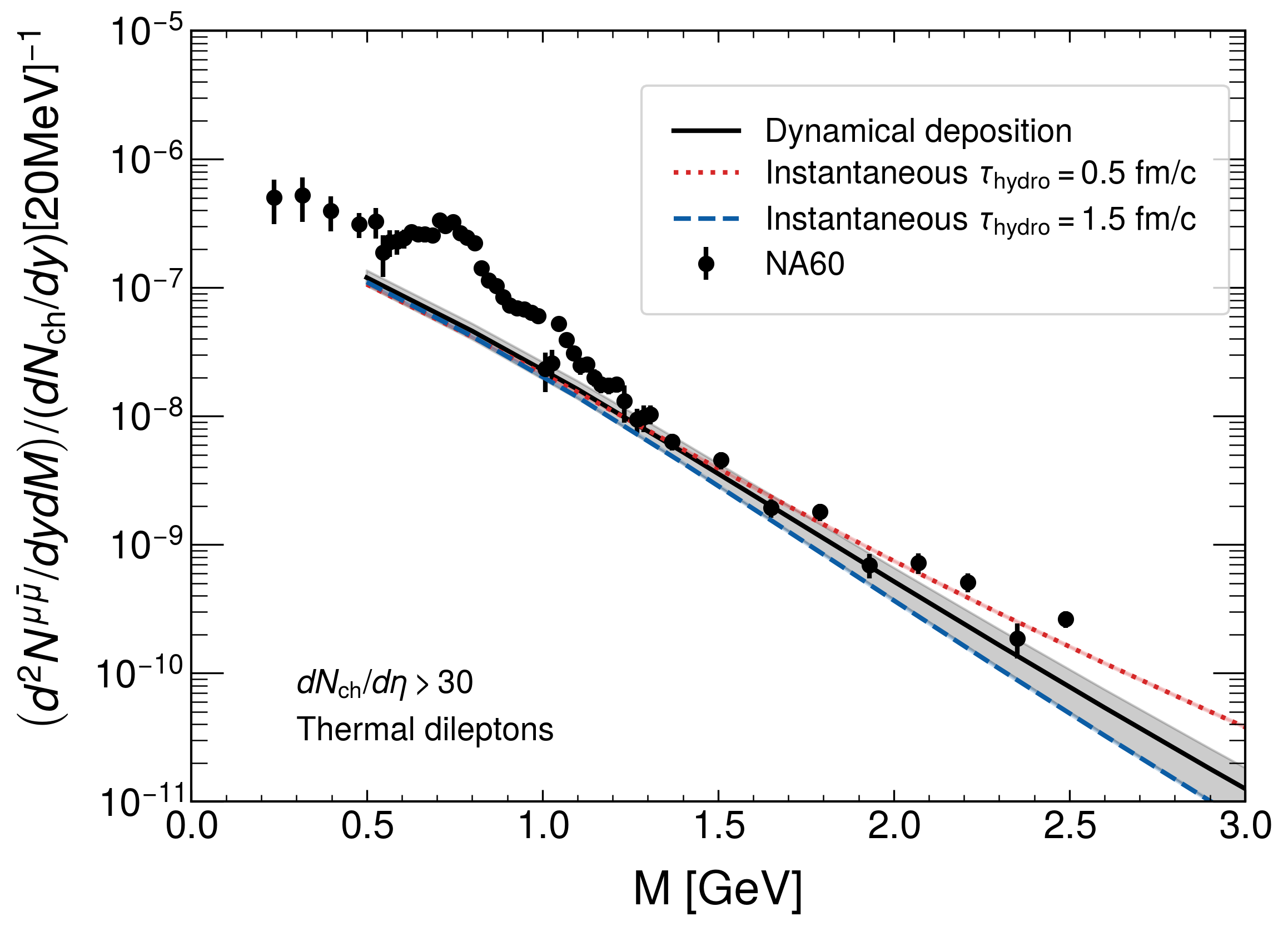 }
 \caption{Dilepton invariant mass excess spectra in In+In collisions for the $dN_{\text{ch}}/d\eta > 30$ centrality class  for the dynamical deposition case and the instantaneous deposition cases with initial proper times $\tau_0$ = 0.5 fm/$c$ and 1.5 fm/$c$. Data are  from the NA60 Collaboration \cite{NA60:2008dcb}. }
	\label{fig:Thermaldilepton_SPS_tau}
\end{figure}

\begin{figure}[tb]
	\centering
	\includegraphics[width=1.0\linewidth]{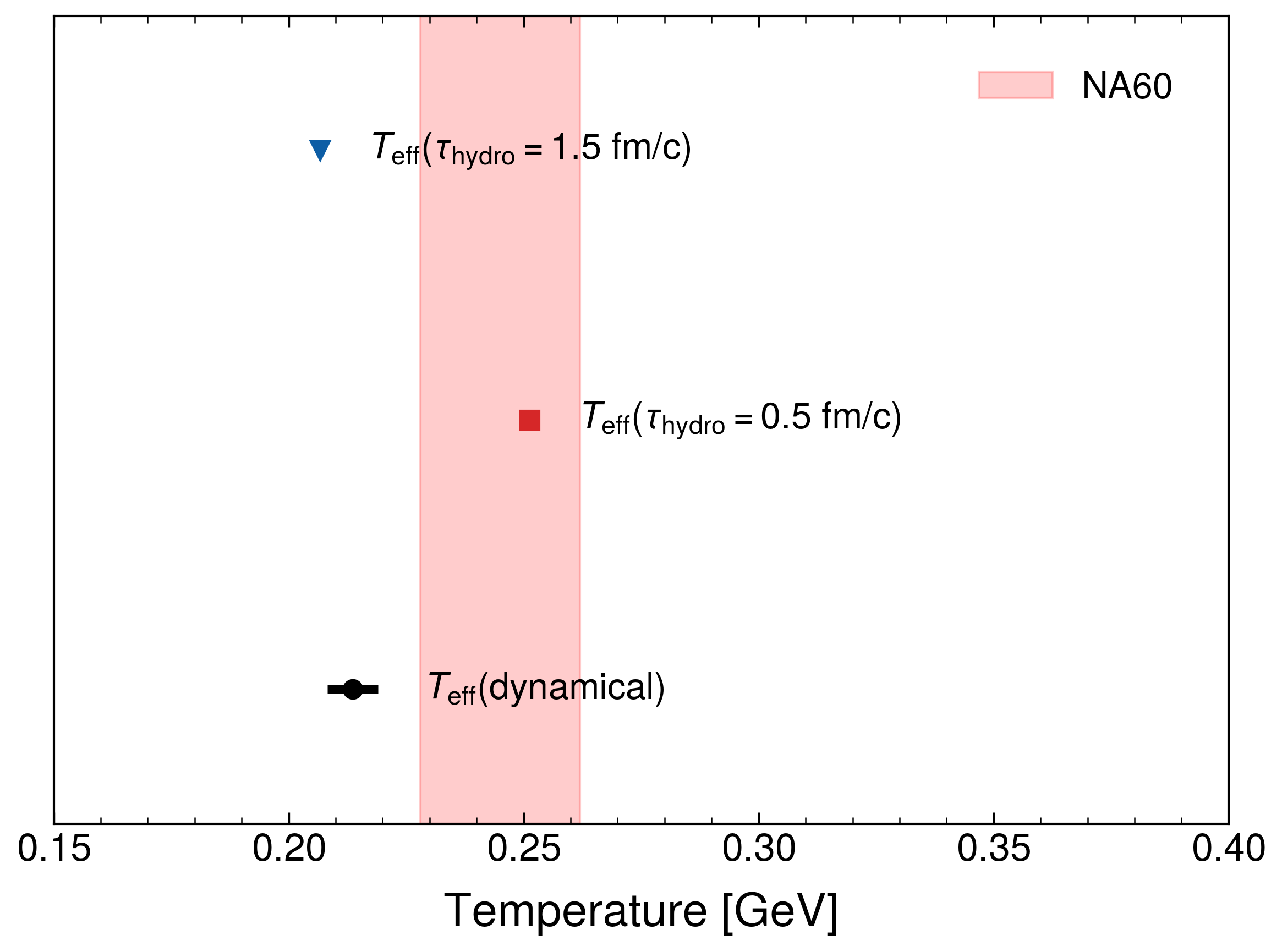 }
 \caption{ Effective temperature ($T_{\rm{eff}}$) extracted from dilepton invariant mass excess spectra in In+In collisions for  $dN_{\text{ch}}/d\eta > 30$, shown for the dynamical deposition case and the instantaneous deposition cases with initial proper times $\tau_0 = $ 0.5 fm/$c$ and 1.5 fm/$c$. The red band represents the data from the NA60 Collaboration \cite{NA60:2008dcb}. }
	\label{fig:Teff_SPS_tau}
\end{figure}

In Fig.~\ref{fig:Thermaldilepton_SPS_tau}, we show the dilepton excess invariant mass spectra for these three scenarios in In+In collisions at $\sqrt{s_{\rm NN}}$ = 17.3 GeV, as measured by the NA60 experiment \cite{NA60:2008dcb}, which provides high-statistics and high-precision dimuon data. The corresponding effective temperatures $T_{\rm eff}$ extracted from these spectra are presented in Fig.~\ref{fig:Teff_SPS_tau}.
The results indicate that the slope of the dilepton spectra and the extracted $T_{\rm eff}$ are sensitive to the pre-equilibrium evolution. Earlier energy deposition leads to a higher $T_{\rm eff}$ and shows better agreement with the data. The dynamical deposition scenario lies between the two instantaneous cases and is closer to the $\tau_0$ = 1.5 fm/$c$ case.
However, it is important to note that at  $\sqrt{s_{\rm NN}}$ = 17.3 GeV collision energy, the $\tau_0$ = 0.5 fm/$c$ initialization time is likely unrealistic because of the longer nuclear crossing time at low collision energies. This implies that dilepton production from the early pre-equilibrium phase, roughly between 0.5 and 1.5 fm/$c$ in this case, may have a sizable contribution to the final dilepton spectra. Therefore, the inclusion of pre-equilibrium dilepton emission is essential for the complete calculation of dilepton spectra, and should be considered in future studies.

A key point that needs to be highlighted is that the uncertainty band shown for the dynamical deposition case in Fig.~\ref{fig:Thermaldilepton_SPS_tau} and Fig.~\ref{fig:Teff_SPS_tau} does not represent statistical uncertainty, but rather arises from sampling the posterior distribution obtained through Bayesian calibration.
This theoretical uncertainty is extracted from the variance difference between the following two sets of simulations. In the first set of simulations, we sample different model parameter sets from the posterior distribution for individual heavy-ion collision events. The variance of the dilepton invariant mass spectrum obtained from this ensemble of simulations includes both event-by-event fluctuations and variations of the model parameters across the entire posterior distribution. To estimate the average variance from the event-by-event fluctuations, we perform a second batch of reference simulations with model parameters fixed to the one with the highest likelihood. Finally, the difference between the variances of the observable from these two sets of simulations represents the theoretical uncertainty associated with the posterior distribution. For further details, see Appendix~\ref{Sec:uncertainty}.

As shown in Fig.~\ref{fig:Thermaldilepton_SPS_tau}, this uncertainty increases with invariant mass. The main reason is that the current posterior distribution is constrained by charged hadron observables, which are mostly sensitive to the later stages of the fireball evolution. In contrast, dileptons in the IMR are emitted predominantly during the early stage of the fireball; the larger $M$, the earlier, which is not perfectly constrained by the present calibration. Therefore, incorporating dilepton observables in the IMR region into future Bayesian analyses may help further constrain the transport properties of the QGP in its early time evolution.

\section{Conclusion}
\label{sec:conclusion}
In this study, we investigated electromagnetic radiation in heavy-ion collisions at RHIC-BES and SPS energies using the iEBE-MUSIC framework. This approach integrates a 3D dynamical Monte Carlo Glauber initial condition, (3+1)D viscous relativistic hydrodynamic evolution (MUSIC), and the UrQMD hadronic afterburner, with model parameters calibrated in a recent Bayesian analysis.
Our framework demonstrates strong predictive power for hadronic observables
of different collision systems, spanning the entire Beam Energy Scan program in the baryon-rich region.

With the calibrated dynamical evolution of the QGP medium, we incorporate 
the thermal photon emission rates in the QGP and hadron gas phases and the prompt photon contribution from NLO pQCD calculations. Using this framework, we systematically study the dependence of direct photon production and their anisotropic flow on different emission sources as a function of collision energy.
At the top RHIC energy, even with the new (3+1)D dynamical initial conditions, the long-standing discrepancy between calculated and measured flow remains. While our results show good agreement with the STAR preliminary direct photon spectra, the photon spectra measured by the PHENIX Collaboration, as well as their direct photon elliptic flow data, are underestimated.

At $\sqrt{s_\mathrm{NN}} \sim O(10)$ GeV with the scaling factor of prompt photon fixed at $\lambda=0.5$, thermal photons with transverse momenta $p_T \in [2.5, 4]$ GeV may outshine the prompt source and lead to sizable flow anisotropies in the direct photons. 
The integrated direct-photon yield exhibits an approximately universal multiplicity scaling, with a scaling exponent $\alpha$ closer to that of the preliminary STAR measurement. These conclusions, however, require further validation once the full direct photon uncertainties are considered. We find that the total theoretical uncertainty is dominated by the contribution of prompt photons. 
Therefore, the treatment of prompt-photon uncertainties requires careful consideration in future theoretical and experimental work.
If future experiments are able to measure prompt photon production in the low-$p_T$ region at low collision energies in pp collisions, the associated uncertainties in current theoretical calculations could be significantly reduced. This, in turn, would enable a more precise study of the QGP’s thermal properties using thermal photons.

Using NLO thermal QCD dilepton emission rates, we also calculated dilepton excess invariant mass spectra in the finite baryon density regions and extracted the effective temperature from the IMR dilepton spectra. 
Within experimental uncertainties, our calculations can describe the data, but slightly underestimate it for some collision energies.
To a large extent, our multimessenger analysis further supports the conclusion that our model can reproduce the space-time dynamic evolution of the plasma. Since dileptons act as a good thermometer for the QGP medium, 
we further extract the effective temperature $T_{\rm eff}$ from the excess invariant mass spectra in the IMR. The extracted $T_{\rm eff}$ shows a decreasing trend with decreasing collision energy. While the results are within the experimental error bars, they tend to sit at the lower bound of the measured uncertainties. 
This comparison illustrates that $T_{\rm eff}$ provides a quantitative description of the spectral slope, which helps test the consistency between theoretical models and experimental data.
Taken together, the results indicate that the current simulation captures the essential features of the space time evolution of the QGP medium, but there is still room for improvement on the theoretical side. In particular, future dilepton studies will further investigate the contributions from out-of-equilibrium corrections, the hadronic phase, and pre-equilibrium dynamics.
To effectively assess the effects of the pre-equilibrium stage, we varied the initial proper time in the instantaneous deposition scenario and compared the results to the default dynamical deposition case. 
We found that the dilepton spectra are sensitive to the evolution of the pre-equilibrium stage, highlighting their potential for constraining early-time dynamics. 

Finally, we explored the extraction of Bayesian posterior uncertainties from the dilepton spectra in In+In collisions measured by NA60 at the SPS.
We found that the uncertainty increases with invariant mass, mainly because the current Bayesian analysis is constrained by hadronic observables.
Overall, our work shows that the multimessenger approach to heavy-ion collisions previously used at higher energy  is amenable to baryon-rich environments at lower temperatures. 
This paves the way for more comprehensive studies of the QCD phase diagram and for more precise extraction of transport coefficients. 
For instance, this framework can and will be used to make predictions for the upcoming EM measurements at NA60+~\cite{NA60DiCE:2025qra}, NICA~\cite{Sissakian:2009zza}, and ALICE Run 3~\cite{ALICE:2022wwr}, and therefore to deepen our understanding of the properties of strongly interacting nuclear matter over a wide range of temperatures and baryon chemical potentials.

\begin{acknowledgments}
This work is supported in part by the Natural Sciences and Engineering Research Council of Canada  (NSERC) [SAPIN-2020-00048 and SAPIN-2024-00026], in part by US National Science Foundation (NSF) under grant number OAC-2004571, and in part by the U.S. Department of Energy, Office of Science, Office of Nuclear Physics, under DOE Award No.~DE-SC0021969 and DE-SC0024232, under DOE Contract No.~DE-SC0012704 and within the framework of the Saturated Glue (SURGE) Topical Theory Collaboration. C.~S. and J.-F.P. acknowledge a DOE Office of Science Early Career Award (DE-SC0021969 and DE-SC0024347).

Numerical simulations presented in this work were partly performed at the Wayne State Grid and partly at the
B\'eluga and Rorqual supercomputer systems from McGill University, managed by Calcul Qu\'ebec and by the Digital Research Alliance of Canada. 
This research was done also using resources provided by the Open Science Grid (OSG) \cite{Pordes:2007zzb, Sfiligoi:2009cct, osg_2006, osg_2015}, which is supported by the National Science Foundation award \#2030508 and \#1836650. 
\end{acknowledgments}

\section{DATA AVAILABILITY}
The data that support the findings of this article are openly
available~\cite{wu_2026_21083043}

\appendix

\section{Prompt photons}
\label{Sec:promptPhoton}

\begin{figure}[htb]
	\centering
	\includegraphics[width=1.0\linewidth]{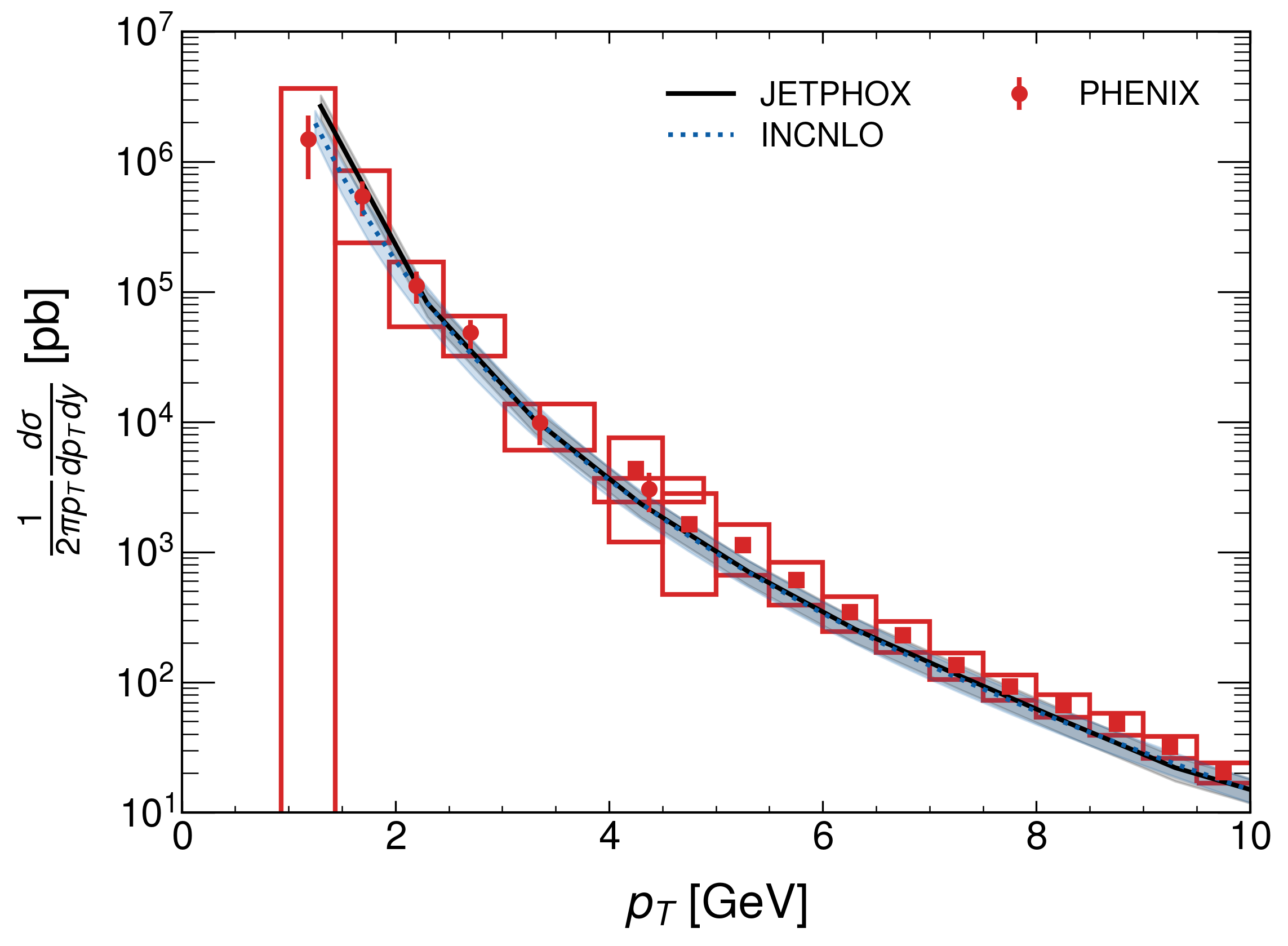 }
 \caption{Transverse momentum ($p_T$) dependent cross section of prompt photon production in $p+p$ collisions at $\sqrt{s} = 200$ GeV, calculated with  JETPHOX and INCNLO. The band represents results obtained when varying the scale factor from $\lambda = 0.5$ to $\lambda = 2.0$. }\label{fig:prompt_photon_pp_JETPHOX}
\end{figure}

In this appendix, we perform an independent check of prompt photon calculations using JETPHOX\cite{Catani:2002ny,Aurenche:2006vj,Belghobsi:2009hx}.
The JETPHOX is widely used for computing prompt photon production in high-energy collisions and is based on NLO pQCD calculations. JETPHOX is flexible in choosing PDFs and isolation criteria.

We use the same setup for JETPHOX, including the nCTEQ15np parton distribution function and BFG-II fragmentation functions. We include both the direct photon contribution\footnote{The direct contribution from $\gamma$+jet production without fragmentation.} and the fragmentation photon contribution\footnote{The contribution from hadron+jet production with one fragmentation photon.} in prompt photon calculations. For JETPHOX, we do not apply any isolation criterion in this comparison. The detailed input file used for JETPHOX in this study can be found in \cite{wu_2025_16929665}.

Figure~\ref{fig:prompt_photon_pp_JETPHOX} shows the $p_T$-dependent cross section of prompt photon production in $p+p$ collisions at $\sqrt{s}=200$ GeV as calculated by JETPHOX and INCNLO packages. The PHENIX data \cite{PHENIX:2012krx} is included as well. The band results from varying the scale $\lambda$ from $0.5$ to $2.0$. 
Clearly, JETPHOX shows good agreement with the PHENIX data and provides results similar to those from INCNLO~\cite{Paquet:2015lta}. 
{This indicates that the model dependence between INCNLO and JETPHOX in prompt photon calculations is minor and can be considered negligible within the current level of uncertainties.}
\section{Thermal Photon Emission rates from QGP 2-to-2 Processes at Finite $\mu_B$}

At $\mu_B=0$, the leading-order soft part of the photon rate takes the form~\cite{Arnold:2001ms,Ghiglieri:2013gia}
\begin{equation}
k \frac{d \Gamma_\gamma}{d^3 k}= \left( \sum_s q_s^2 \right) \frac{\alpha_{EM} m_\infty^2 N_c}{(2\pi)^3}  n_F(k) \ln\left(\frac{\mu_\perp^2}{m_\infty^2}+1\right)
\end{equation}

The factor $n_F(k)$ comes from the relation
\begin{equation}
n_B(k)\left(1-2 n_F(k)\right)=n_F(k)
\end{equation}

At finite $\mu_B$, this $n_F(k)$ factor becomes
\begin{equation}
n_B(k) \frac{1}{2}\left[\left(1-2 n_F(k,\mu_q)\right) + \left(1-2 n_F(k,-\mu_q)\right) \right]
\end{equation}
where $\mu_q=\mu_B/3$ is the quark chemical potential.

The final result,
\begin{align}
&k \frac{d \Gamma_\gamma}{d^3 k}= \left( \sum_s q_s^2 \right) \frac{\alpha_{EM} m_\infty^2 N_c}{(2\pi)^3}  n_B(k)\notag \\
&\hspace{4em}\left[1- n_F(k,\mu_q)-n_F(k,-\mu_q) \right] \ln\left(\frac{\mu_\perp^2}{m_\infty^2}+1\right)
\end{align}
reduces to the result from Ref.~\cite{Traxler:1994hy} in the limit $k\gg T$ and $k \gg \mu_q$.

The leading-order hard part is evaluated numerically as originally described in Ref.~\cite{Arnold:2001ms,Gervais:2012wd}. 

\section{Model Uncertainty Estimated from the Posterior}
\label{Sec:uncertainty}

In this appendix, we make theoretical predictions with uncertainties by varying the model parameters according to the posterior distribution.
The observable $O$ is computed as
\begin{align}
    O = \bar{O} \pm \Delta O
\end{align}
with
\begin{align}
    \bar{O} = \frac{1}{N_\mathrm{params}}  \sum_{i = 1}^{N_\mathrm{params}} \frac{1}{N_\mathrm{ev}} \sum_{j = 1}^{N_\mathrm{ev}} O_j({\bm{\theta}}_i)
    \label{eq:Omean}
\end{align}
and
\begin{align}
    \Delta O = \mathrm{std} \left(\frac{1}{N_\mathrm{ev}} \sum_{j = 1}^{N_\mathrm{ev}} O_j({\bm{\theta}}_i)\right)
    \label{eq:DeltaO}
\end{align}
Here, $O_j(\bm{\theta}_i)$ denotes the result for the $j$-th event when computing the observable $O$ with the $i$-th parameter set drawn from the posterior distribution. The standard deviation is calculated over the $N_{\mathrm{params}}$ parameter sets. In standard Bayesian analysis, \(N_{\mathrm{params}}\) and \(N_{\mathrm{ev}}\) are ideally taken as large as possible to sample many points; however, the computational cost is substantial.

One way to estimate Eq.~\eqref{eq:Omean} is to fix $N_\mathrm{ev}=1$ and take $N_\mathrm{params}$ as large as possible. Because the choice of model parameters from the posterior distribution is independent of simulating a Monte Carlo event, we expect the mean of the observable, computed as
\begin{align}
    \bar{O}_\mathrm{conv} = \frac{1}{N_\mathrm{params}}  \sum_{i = 1}^{N_\mathrm{params}} O_1({\bm{\theta}}_i)
    \label{eq:OmeanAppro1}
\end{align}
 converge to Eq.~\eqref{eq:Omean} when $N_\mathrm{params} \rightarrow \infty$.

From this set of simulations, the variance of the observable  $O$ is computed, which contains fluctuations due to variations in the model parameter sets and the Monte Carlo events
\begin{align}
    \sigma_{O, \mathrm{conv}}^2 = \sigma^2_{O, \mathrm{param}} + \sigma^2_{O, \mathrm{ev}}.
\end{align}

To estimate the theory uncertainty $\Delta O \equiv \sigma_{O, \mathrm{param}}$, another set of events is simulated with the model parameters fixed at the MAP. In this second set, the variance of $O$ only contains the fluctuations from the Monte-Carlo events,
\begin{align}
    \sigma^2_{O(\mathrm{MAP})} = \sigma^2_{O, \mathrm{ev}}.
\end{align}
It is assumed that the variance due to Monte Carlo events is independent of the parameter set $i$, $\sigma^2_{O( \theta_i)} = \sigma^2_{O(\mathrm{MAP})} \equiv \sigma^2_{O,\mathrm{ref}}$. The theory uncertainty is then estimated as
\begin{align}
    \Delta O = \sqrt{\sigma_{O, \mathrm{conv}}^2 - \sigma^2_{O,\mathrm{ref}}}.
\end{align}
It should be noted that the method applies only to simple observables defined as event averages, such as particle spectra.

\section{QGP and hadronic gas contributions in thermal photon emission}
In this appendix, we present thermal photon (QGP+HG) observables and their separate contributions from the QGP and hadronic gas (HG) phases. Following the setup in Sec. \ref{sec:EM_sources}, thermal photons from the QGP phase are calculated using fluid cells with $T_{\rm{tr}} > 0.16$ GeV, using leading-order rates with shear and bulk viscous corrections. The HG phase contribution includes fluid cells with $0.12 < T_{\rm{tr}} < 0.16$ GeV, incorporating all hadronic processes described in Sec.~\ref{sec:EM_sources}.

\begin{figure}[htb]
	\centering
	\includegraphics[width=1.0\linewidth]{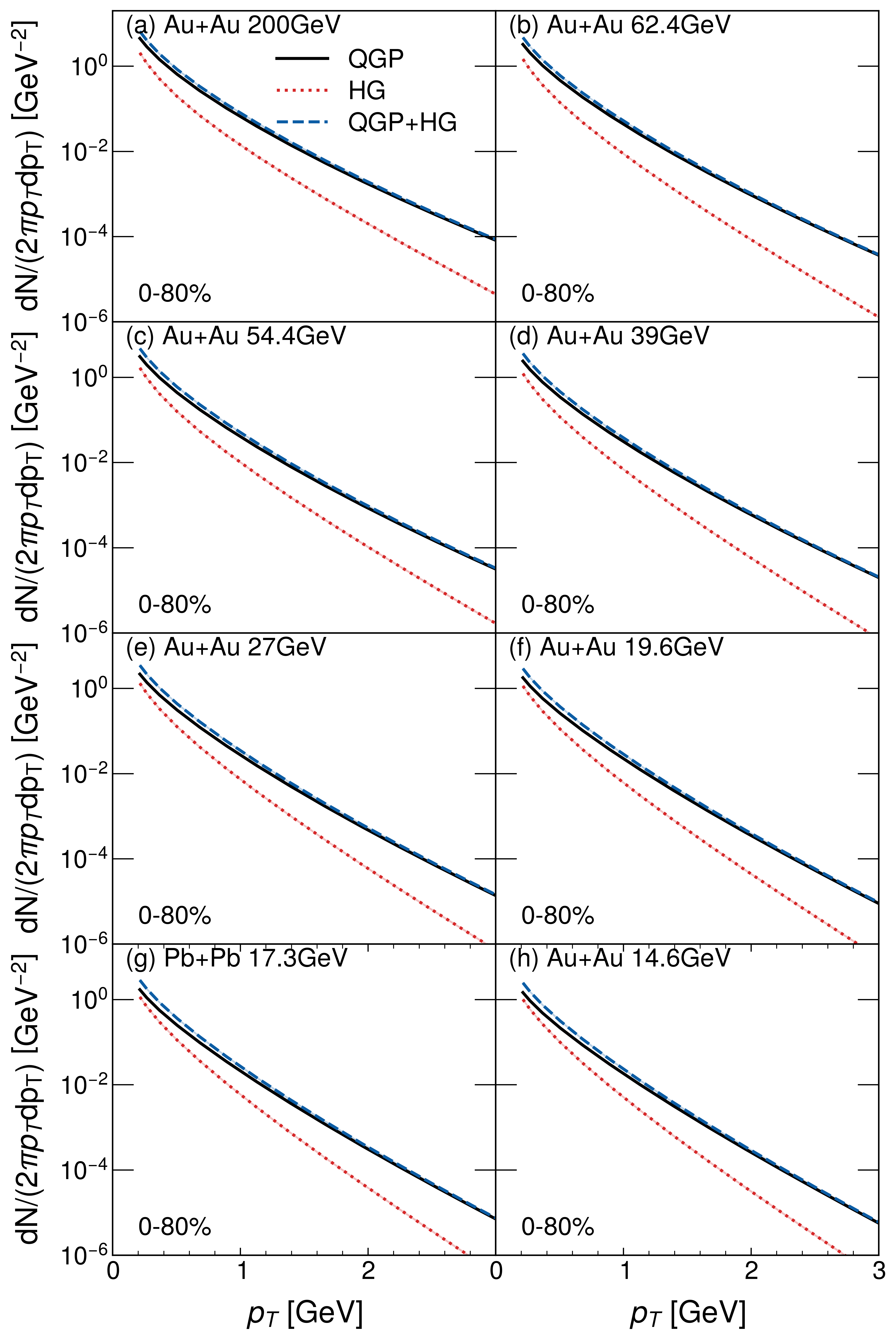 }
 \caption{Transverse momentum ($p_T$) distributions of thermal photon production (QGP+HG) and the individual contributions from QGP and HG phases in minimum-bias Au+Au/Pb+Pb collisions at RHIC-BES/WA98 energies.
}\label{fig:dNdpT_QGP_HG}
\end{figure}
Similar to Fig.~\ref{fig:photons_RHICBES}, we plot the $p_T$ distribution of thermal photon production and its separate contributions from the QGP and HG phases in minimum-bias Au+Au/Pb+Pb collisions at RHIC-BES and WA98 energies in Fig.~\ref{fig:dNdpT_QGP_HG}. From $\sqrt{s_{\rm{NN}}} = 14$ to 200 GeV, the thermal photon yield is dominated by the QGP phase across a wide $p_T$ range. 
The QGP contribution decreases toward lower collision energies and lower $p_T$. Nevertheless, even at $\sqrt{s_{\rm{NN}}} = 14.6$ GeV and $p_T \sim 0.2$ GeV, it still accounts for more than 50\% of the total yield. This is evident from the hydrodynamic expansion trajectories in Fig.~\ref{fig:trajectories} for a single Au+Au event with $b = 0$ and $|\eta_s| < 1$, which show that the system still spends a substantial part of its evolution in the QGP phase. In addition, the slower energy deposition at lower collision energies also reshuffles the temperature range over which thermal photons are emitted, further reinforcing the dominant contribution of QGP photons to the total thermal spectrum.

\begin{figure}[htb]
	\centering
	\includegraphics[width=1.0\linewidth]{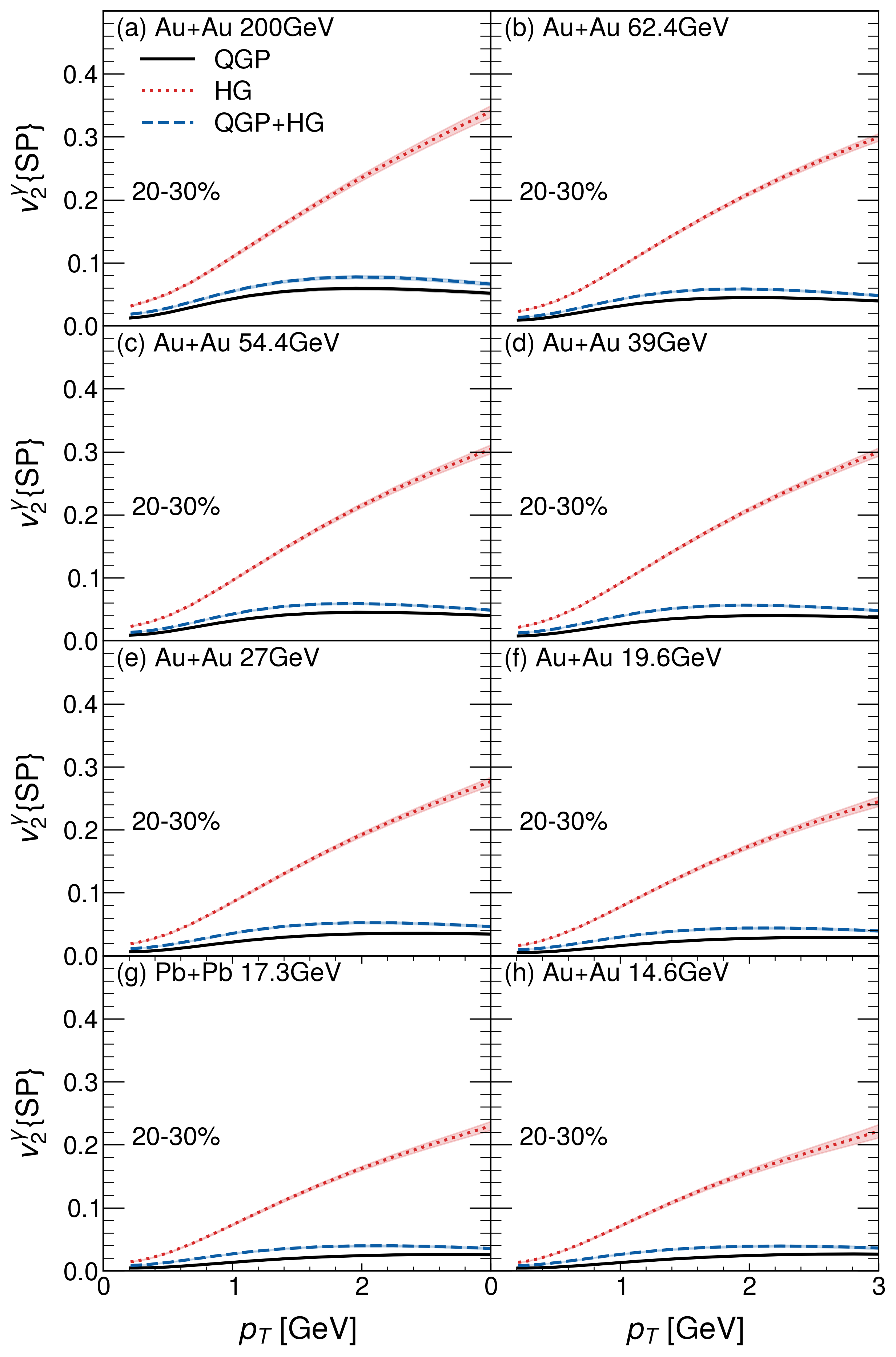 }
 \caption{Thermal photon elliptic flow ($v_2\{\mathrm{SP}\}$) with the individual contributions from QGP and HG phases in the 20–30\% centrality class for Au+Au collisions at $\sqrt{s_{\rm NN}}$ = 14.6 -- 200 GeV, and for Pb+Pb collisions at $\sqrt{s_{\rm NN}} = 17.3$ GeV.
}\label{fig:v2_QGP_HG}
\end{figure}
In Fig.~\ref{fig:v2_QGP_HG}, we also present the elliptic flow ($v_2\{\rm SP\}$) of thermal photons and their individual contributions from the QGP and HG phases in the 20–30\% centrality class, for Au+Au collisions at $\sqrt{s_{\rm NN}} = 14.6$ -- 200 GeV and for Pb+Pb collisions at $\sqrt{s_{\rm NN}} = 17.3$ GeV. Photons from the HG phase can produce a sizable $v_2$ because they are emitted from lower-temperature regions and the later stages of the system evolution. However, as shown in Fig.~\ref{fig:dNdpT_QGP_HG}, their overall contribution to the total thermal photon yield is subleading. As a result, the final $v_2$ of thermal photons is largely dominated by the QGP phase.

\begin{figure}[htb]
	\centering
	\includegraphics[width=1.0\linewidth]{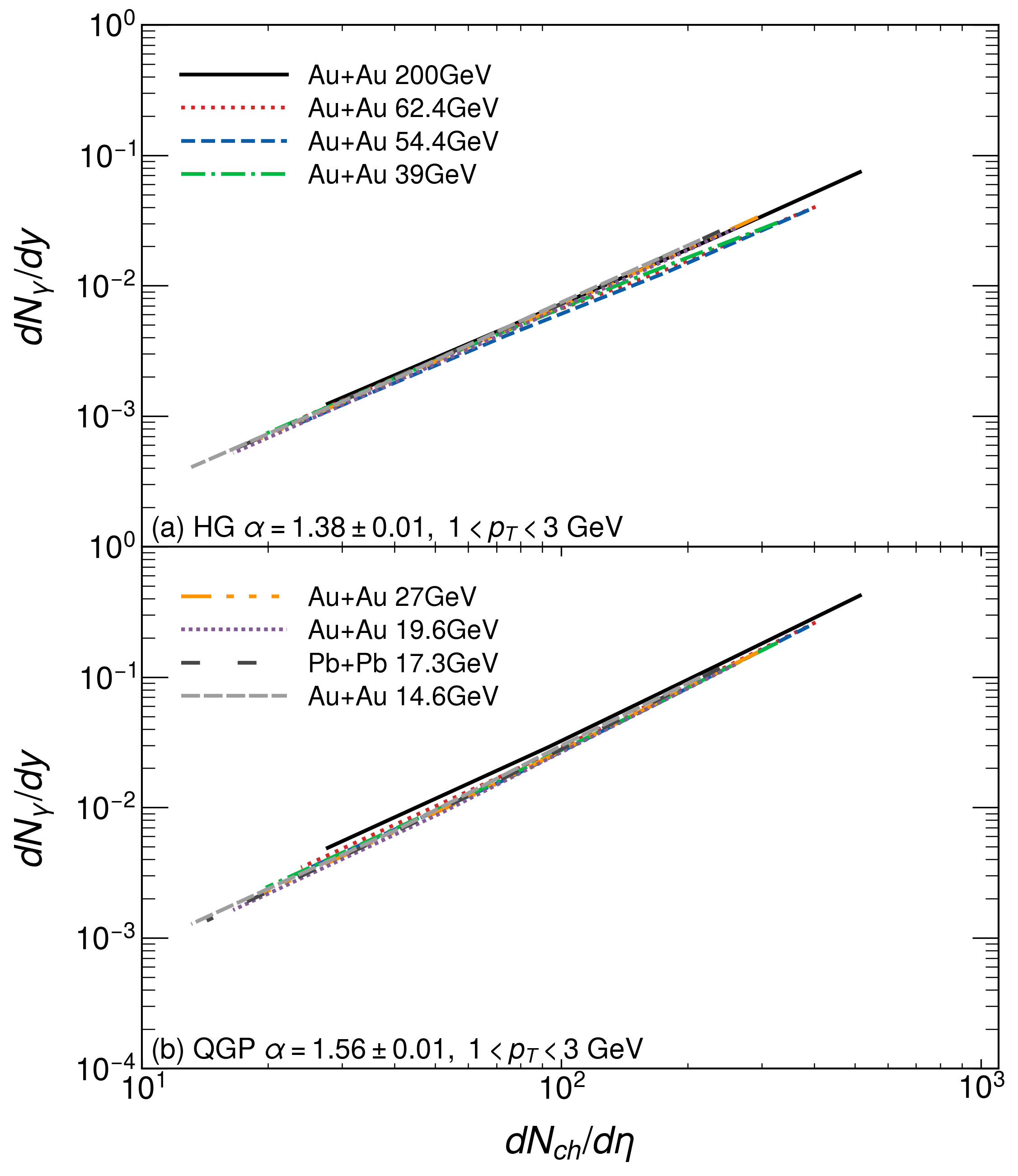 }
 \caption{Multiplicity ($dN_{\rm{ch}}/d\eta$) dependence of integrated thermal photon yields from the QGP phase (panel a) and the HG phase (panel b). 
}\label{fig:dNgammady_scaling_HG_QGP}
\end{figure}
Finally, Fig.~\ref{fig:dNgammady_scaling_HG_QGP} shows the dependence of integrated thermal photon yields on charged particle multiplicity ($dN_{\mathrm{ch}}/d\eta$), separated into QGP and HG components. It is found that the thermal photon yields from both the QGP and HG approximately follow a power-law $(dN_{\mathrm{ch}}/d\eta)^\alpha$ curve, similar to the total thermal photon results shown in Fig.~\ref{fig:dNgammady_scaling}. The QGP component exhibits a larger exponent $\alpha$ than that of the HG phase, and is close to the exponent extracted from the total thermal photon yield. Similar results have been reported in \cite{Shen:2013vja}.

\bibliography{inspire, non-inspire}

\end{document}